\documentclass[showpacs,preprint,amsmath,amssymb]{revtex4}

\usepackage[dvips]{graphicx}   
\usepackage{dcolumn}           
\usepackage{bm}                

\begin{document}
\title{Chain morphology, swelling exponent, persistence length, like-charge attraction,
and charge distribution around a chain in polyelectrolyte solutions:
effects of salt concentration and ion size studied by molecular dynamics simulations}
\author{Pai-Yi Hsiao} 
\email{pyhsiao@ess.nthu.edu.tw}
\affiliation{Department of Engineering and System Science, 
National Tsing Hua University, Hsinchu, Taiwan 300, R.O.C.}

\date{\today} 

\begin{abstract} 
Properties of polyelectrolytes in tetravalent salt solutions are intensively
investigated by a coarse-grained model. 
The concentration of salt and the size of tetravalent counterions are found 
playing a decisive role on chain properties. 
If the size of tetravalent counterions is compatible with the one of monomers,
the chains show extended structures at low and at high salt concentrations,
whereas at intermediate salt concentrations, they acquire compact and 
prolate structures.
The swelling exponent of a chain against salt concentration behaves in an 
analogous way as the morphological quantities. 
Under certain condition, the electrostatics gives a negative contribution 
to the persistence length, in companion with a salt-induced mechanical 
instability of polyelectrolytes.  Nearly at the same moment, it appears 
like-charge attraction between chains.  The equal size of the tetravalent 
ions and the monomers is the optimal condition to attain the strongest 
attraction between chains and the most compact chain structure. 
Moreover, the ions form a multi-layer organization around a chain
and, thus, the integrated charge distribution reveals an oscillatory behavior.
The results suggest that charge inversion has no direct connection 
with redissolution of polyelectrolytes at high salt concentrations.

\end{abstract} 


\maketitle

\section{Introduction} 
Solutions of highly-charged polyelectrolytes, such as biopolymers (DNA, proteins, \textit{etc.}) 
and many synthetic polyelectrolytes, exhibit striking phenomena while multivalent salts or 
charged molecules are added into the solutions~\cite{bungenbergdejong31,ikegami62,delacruz95,raspaud98}.
If the concentration of the added salt lies within a window of concentration, 
the solution is separated in two phases: 
one is a viscous phase containing condensed or precipitated polyelectrolytes, 
and the other is a dilute phase in which polyelectrolytes are dissolved.  
If it lies outside the window, the solution is in a homogeneous phase. 
These phenomena are called ``reentrant condensation''~\cite{nguyen00}, 
or commonly referred as ``DNA condensation'' in the fields of biology and medicine~\cite{bloomfield96}. 
This topic has attracted much attention recently due to its potential application
in developing nonviral vectors of DNA for the purpose of gene 
therapy~\cite{vijayanathan02}.  
The amount of salt which results in the phase separation is determined by many factors,
such as valence of salt, temperature, chain structure, and concentration of 
polyelectrolytes~\cite{kruyt49,ikegami62,delacruz95,raspaud98}. 
It is known that the lower boundary of the window depends linearly on the 
concentration of polyelectrolytes but the upper one of the window is insensitive 
to it and roughly a constant.  
Both flexible polyelectrolytes and stiff polyelectrolytes display similar
phase diagrams~\cite{delacruz95,raspaud98,saminathan99}.

Despite decades of effort, our understanding on polyelectrolytes is still limited 
and many issues, including the reentrant condensation,  are far from being 
resolved~\cite{tripathy02}.  
The aim of this paper is to investigate the properties of polyelectrolytes in salt
solutions by means of computer simulations. 
There have been many simulations devoted to the study of 
polyelectrolytes~\cite{stevens99,stevens01,chang03b,stevens95,winkler98,liu02,stevens98,liu03,klos05a,khan05,dias03,sarraguca03,deserno02}.
However, most of them investigated the properties in salt-free 
solutions~\cite{stevens99,stevens01,chang03b,stevens95,winkler98,liu02}.
Only recently, due to the progress of computing power, large-scale simulations
for systems with added salt can be realized.  
The works included, for example, flexible chains in monovalent or multivalent 
salt solutions~\cite{stevens98,liu03,klos05a,khan05}, semiflexible chains with 
multivalent salt or charged molecules~\cite{dias03,sarraguca03}, 
and rigid chains in the presence of multivalent salts~\cite{deserno02}.  
But the salt concentration or the salt valence was not large enough to observe 
any significant evidence being able to connect to the redissolution of 
polyelectrolytes.  
Therefore, the reentrant condensation could not be discussed thoroughly
in these works.
Allahyrarov \textit{et al.}~\cite{allahyarov04} recently studied the effective interaction
between two immobile DNAs in the presence of tetravalent counterions.
They have found that the effective interaction yields an attractive minimum at
short distances which then disappears in favor of a shallow minimum at large
separations, indicating the occurrence of DNA condensation and redissolution and
a novel stable mesocrystal.  In a more recent study~\cite{hsiao06a}, we have 
investigated the properties of mobile polyelectrolytes in tetravalent salt
solutions and demonstrated that 
following a collapsed transition, polyelectrolytes undergo reexpanding
transition upon addition of the salt.  This phenomenon is just a single-chain
version of the condensation and the redissolution of polyelectrolytes.
Moreover, by varying the ion size, we have shown that the excluded volume 
of ions plays a decisive role on the reentrant condensation.
However, to well understand the phenomena, there is still a far way to go.  For
example, is it possible to obtain any information related to the phase 
boundary of reentrant condensation from simulations?  There exist theories to 
explain the phase boundary~\cite{delacruz95,raspaud98,nguyen00} but
it has never been studied in simulations.  Concerning the size of
polyelectrolyte in multivalent salts, Muthukumar and coworkers have applied
single and double screening theories to explain the collapsed transition of
chains~\cite{liu03,muthukumar87,muthukumar96}.  However, no existing theory can 
predict chain reexpansion in the region of high salt concentration. Moreover,
chain morphology is essential to understand the properties of polyelectrolytes.  
It can be described by many quantities, such as asphericity and
prolateness parameter~\cite{christos89}.  To our knowledge, no such study 
has been reported yet for strongly-charged polyelectrolytes. 
An alternative way to understand polymers is to study scaling behavior.
Although the concept of scaling has been successfully applied to neutral
polymers~\cite{degennes79}, application of this concept to
polyelectrolytes~\cite{peterson96,schiessel98,schiessel99,liao03} has
encountered many difficulties owing to the long-range nature of
Coulomb interaction.  In this paper, we compute the single-chain structure
factor.  The swelling exponent is then calculated by studying
the power-law-like regime of the structure factor.  
Since properties of polyelectrolytes depend strongly on ion size, we investigate
in this paper the effect of ion size too.  In our previous work~\cite{hsiao06a}, the sizes of ions 
were set to identical and varied simultaneously.  As a consequence,
Bjerrum association~\cite{bjerrum26} was strongly intervened, particularly when
ions were small.  This setup has a disadvantage that the system was easily jammed
by large ions, leading to the impossibility of investigation of the cases of big ions.
Therefore,  in this study we vary only the size of the multivalent
counterions and fix unchanged the one of the other ions, so that the ion
size effect is restrictedly investigated. 

In the study of polyelectrolytes, it is tempting to discuss the long range
nature of Coulomb interaction using the concept of electrostatic persistence
length.  Odijk~\cite{odijk77}, Skolnick and Fixman~\cite{skolnick77} 
predicted that the electrostatic persistence length is proportional to the
square of the screening length $r_{\rm s}$.  The deduction was based upon
Debye-H\"uckel theory. Consequently, the effective intramolecular
interaction is always repulsive, resulting in a positive electrostatic
persistence length~\cite{lebret82,fixman82}.  However, experiments have shown
that electrostatic persistence length can be negative under certain condition
so that electrostatics gives a reversed contribution to chain  
flexibility~\cite{baumann97,hugel01}.  Simulations on this 
issue~\cite{micka96,ullner97,nguyen02,everaers02,ullner03} were mainly
performed within Debye-H\"uckel approximation using a screened Coulomb potential
$U_{\rm DH}(r)\propto e^{-r/r_{\rm s}}/r$ to describe the screening effect of 
salt without involving explicitly salt ions.
A negative electrostatic persistence length, hence, has never been reproduced
in the simulations.  Noticeably, it is known that linear screening theory 
is valid only for a weakly-charged system and breaks down for a highly-charged one, 
such as DNA and many biopolymers, due to strong ion 
correlations~\cite{guldbrand84,stevens96,stevens98}.  
Recently, by employing a loop expansion method which goes beyond mean field, 
Ariel and Andelman were able to predict the negative regime of the electrostatic 
persistence length for a strongly-charged rodlike polyelectrolyte~\cite{ariel03b}.  
Therefore, we expect that to observe a negative electrostatic persistence length, 
one should incorporate salt ions explicitly and use the bare 
Coulomb potential in simulations.
Our study fulfills this requirement and hence offer a
good chance to verify the above phenomenon and theory.   

Charge inversion is a universal phenomenon, occurred when a strongly charged
macroion binds so many counterions that its net charge alters
sign~\cite{grosberg02}.  Nguyen \textit{et al.} intended to link charge
inversion with the reentrant condensation of polyelectrolytes~\cite{nguyen00}.
The idea was based upon that counterions form strongly correlated liquid on the
surface of polyelectrolytes. 
In an intermediate salt region, the bare charges of the chains are almost neutralized 
by the condensed counterions; the correlation-induced attraction results in the 
condensation of polyelectrolytes.
In a high-salt region, the counterions overcompensate the chain bare charges;
Coulomb repulsion between chains dominates the attraction and, in consequence, the condensates 
are redissolved into the solution. 
Solis and Olvera de la Cruz gave a different point
of view using a theory called ``two-state model''~\cite{solis00,solis01}.  They predicted that 
charge inversion is not necessarily occurred with redissolution of polyelectrolytes.  
Although overcompensated by counterions, the redissolved polyelectrolytes can have
either sign due to the association of coions~\cite{solis02}.  This
association depends strongly on the ion size which was not considered
in the study of Nguyen \textit{et al.}~\cite{nguyen00}.  Therefore, Solis and Olvera de
la Cruz predicted a charge distribution which oscillates from a
polyelectrolyte, in accordance with the profile observed in our previous
work~\cite{hsiao06a}.  In this paper, we investigate more extreme cases, including
much smaller and much larger ions.  We  also study like-charge attraction between 
chains by calculating the potential of mean force and determine the conditions
under which the attraction takes place.

The rest of the article is organized as follows.  Model and simulation details
are given in Sec.~\ref{Sec_Model_Method}.  Results are discussed in
Sec.~\ref{Sec_Result_Discussion}. It is divided into five subsections.  The
first subsection (Sec.~\ref{Sec_Rg2_Cm}) is devoted to the study of the radius
of gyration of a polyelectrolyte where salt concentration and polyelectrolyte
concentration are varied.  The results provide information helpful in understanding 
the boundaries of the condensation window.  Sec.~\ref{Sec_Shape} studies the
morphology of multiple polyelectrolytes in salt solutions.  We compute several
quantities, including shape factors, asphericity, and prolateness parameter.
The effect of size of multivalent ions is discussed.  Single-chain structure
factor is, then, calculated in Sec.~\ref{Sec_SF}.  By least-square fitting the
power-law-like regime, we compute the swelling exponent of a polyelectrolyte,
which provides a fundamental vision how strongly-charged polyelectrolytes scale
in salt solutions.  Sec.~\ref{Sec_lp} deals with
persistence length.  The negative regime of the electrostatic persistence
length is successfully reproduced in the simulations. Comparison of the results
with theoretical predictions is made.  Sec.~\ref{Sec_pmf} provides a
general picture of the effective interaction between polyelectrolytes by computing
the potential of mean force.  As long as knowing the condition to appear
like-charge attraction between chains and the integrated charge distribution
around a chain (Sec.~\ref{Sec_IonDist}), we are able to arbitrate if the
phenomena of charge inversion and the redissolution of polyelectrolytes are
related.  We give our conclusions in Sec.~\ref{Sec_Conclusion}.

\section{Model and simulation method} 
\label{Sec_Model_Method} 
We employed a coarse-grained model to simulate a polyelectrolyte system.  The
system contains four bead-spring chains, one of which is composed of 48 beads
(monomers).  Each monomer dissociates a monovalent cation (counterion) into a solution and
carries a negative unit charge.  There are totally 192 monovalent cations in
the solution.  Solvent molecules are not modeled explicitly.  Their effect is
taken into account implicitly by considering them as a medium of  constant
dielectric constant.  The geometry of the medium is cubic and periodic boundary
condition is applied to simulate bulk environment.  Salt molecules are added
into the system.  In the solution, they are dissociated into cations (counterions) and
anions (coions).  We assume that the cations are tetravalent and the anions are monovalent.  
The amount of the cations and anions obeys
the condition of charge neutrality.

Three types of interactions are included in our simulations. 
The first one is the excluded volume interaction described 
by a purely repulsive Lennard-Jones (LJ) potential, 
\begin{equation}
U_{\rm LJ}(r)=
\left\{ \begin{array}{ll} 
4\varepsilon_{\rm LJ}\left[ \left(\sigma/r\right)^{12}
-\left(\sigma/r\right)^{6}\right] + \varepsilon_{\rm LJ}
& \mbox{\ for $r \le \root 6 \of 2 \sigma$\;}\\
0 & \mbox{\ for $r > \root 6 \of 2 \sigma$\;}
\end{array}
\right.,
\end{equation}
where $r$ is the distance between two particles, $\varepsilon_{\rm LJ}$ is
the interaction strength, and $\sigma$ is the collision diameter. 
We assume that all the particles have identical $\varepsilon_{\rm LJ}$
but the collision diameter may be different.
The collision diameters for monomer, monovalent cation,
tetravalent cation, and anion are denoted by $\sigma_{\rm m}$, 
$\sigma_{+1}$, $\sigma_{+4}$, and $\sigma_{-1}$, respectively.
Lorentz-Berthelot mixing rule~\cite{allen87book} is applied for 
the interaction between different kinds of particles.

The second interaction is called finitely extensible nonlinear elastic 
(FENE) potential~\cite{dunweg91} and used to describe the bond connection 
between adjacent monomers on a chain.  It reads as
\begin{equation}
U_{\rm FENE}(\ell)= -\frac{1}{2} k_{\rm FENE} R_0^2\;
\ln\left( 1- \frac{\ell^2}{R_0^2} \right)\;,
\end{equation}
where $k_{\rm FENE}$ is the spring constant, $\ell$ is bond length, 
and $R_0$ is the maximum extension of a bond.
All particles interact via the third interaction, 
Coulomb interaction, which is  written as
\begin{equation}
U_{\rm coul}(r) = k_{\rm B} T \ell_{\rm B} \frac{Z_i Z_j}{r}
\end{equation}
where  
$Z_i e$ is the charge  of the $i$th particle, 
$T$ is the temperature, $k_{\rm B}$ is the Boltzmann constant, 
and $\ell_{\rm B}$ is the Bjerrum length
defined as the separation distance at which the Coulomb potential 
between two unit charges $e$ equals to the thermal energy $k_{\rm B} T$.
$\ell_{\rm B}$ is equal to  
$e^2/(4\pi\epsilon\epsilon_0k_{\rm B} T)$  
where $\epsilon$ is the dielectric constant and 
$\epsilon_0$ is the vacuum permittivity. 

We perform molecular dynamics (MD) simulations in this study~\footnote{This
work was performed using molecular dynamics simulator LAMMPS\@. 
For more information about LAMMPS, refer to 
http://www.cs.sandia.gov/\ensuremath{\sim}sjplimp/lammps.html.}.  We set
$\sigma_{+1}$ and $\sigma_{-1}$ equal to $\sigma_{\rm m}$ and vary the
size of tetravalent cation $\sigma_{+4}$ from $0.0\sigma_{\rm m}$ to
$4.0\sigma_{\rm m}$ to investigate the effect of ion size on properties of the
polyelectrolyte solution.  Simulation parameters are chosen as follows:
$\varepsilon_{\rm LJ}=0.8333 k_{\rm B} T$, $k_{\rm FENE}=5.8333k_{\rm B}
T/\sigma_{\rm m}^2$, and $R_0= 2 \sigma_{\rm m}$.  The Bjerrum length is set to
$\ell_{\rm B}=3\sigma_{\rm m}$.  Similar systems in salt-free solutions have
been investigated by Stevens and Kremer~\cite{stevens95}.  We assume that the
masses of all the particles are identical and equal to $m$, and choose
$\sigma_{\rm m}$ and $k_{\rm B}T$ to be the length unit and the energy unit,
respectively.  Therefore, the natural time unit is $\tau=\sigma_{\rm
m}\sqrt{m/(k_{\rm B}T)}$.  The system is studied in canonical ensemble where
the temperature is controlled using Nos\'e-Hoover thermostat with the frequency
of heat transfer with thermostat equal to $1.0\tau^{-1}$.  Coulomb interaction
is calculated by the technique of Ewald sum.  Except in Sec.~\ref{Sec_Rg2_Cm},
the monomer concentration is fixed at $C_{\rm m}=0.008 \sigma_{\rm m}^{-3}$
which corresponds to a box size $L=28.8\sigma_{\rm m}$, whereas the salt concentration
is varied.  The amount of salt is controlled such that the volume fraction
of total particles is smaller than $30\%$.  Equations of motion are integrated
applying Verlet algorithm where time step is set to $0.005 \tau$.
Each simulation starts with an equilibration phase of $2 \times 10^6$ MD steps
and is followed by a production run of $10^7$ MD steps in which data are
collected every 1000 steps. 
In order to shorten the notation, we will use $\sigma_{\rm m}$ as the length unit,
$\sigma_{\rm m}^{-3}$ as the concentration unit , and $e$ as the charge unit in 
the following text.

\section{Results and discussions}
\label{Sec_Result_Discussion}
\subsection{Radius of gyration of a polyelectrolyte at different monomer concentrations} 
\label{Sec_Rg2_Cm}
\begin{figure}[htbp] 
\begin{center}
\includegraphics[width=0.4\textwidth,angle=270]{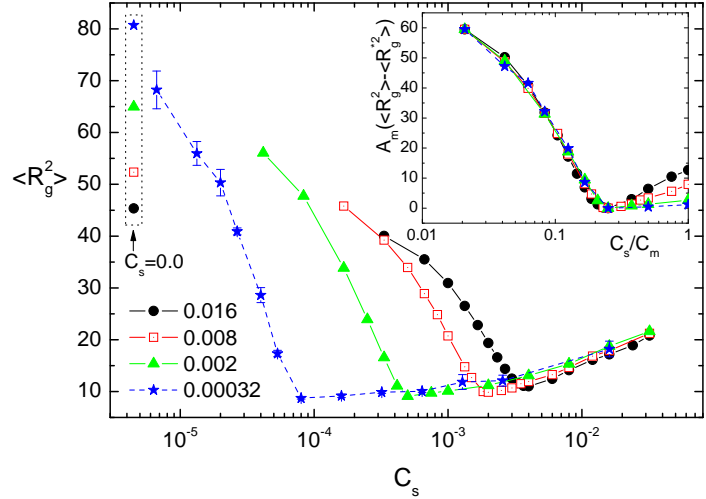}
\caption{$\langle R_{\rm g}^2 \rangle$ as a function of $C_{\rm s}$ at
four monomer concentrations, $C_{\rm m}=0.00032$, $0.002$, $0.008$,
and $0.016$.
The symbols used to represent the four $C_{\rm m}$ are indicated 
at the left-bottom side of the figure.
The value of $\langle R_{\rm g}^2 \rangle$ in a salt-free solution 
($C_{\rm s}$=0.0) is depicted near the left vertical axis.  
The four curves in the regime $C_{\rm s} \le C_{\rm s}^{*}$
can overlap each other (shown in the inset)
under a suitable transformation described in the text.  }
\label{Cm_P1N48_Rg2.eps}
\end{center}
\end{figure}

We started our study by simulating a single polyelectrolyte with added tetravalent salt 
at various monomer concentrations.
The chain consists of 48 monomers and the size of the tetravalent cations
is set to be identical to the one of the monomers. 
We used the radius of gyration to characterize the chain size.
The radius of gyration $R_{\rm g}$ is calculated by the equation
\begin{equation}
R_{\rm g}^2=\frac{1}{N}\sum_{i=1}^{N} (\vec{r}_i-\vec{r}_{\rm cm})^2
\end{equation}
where $N$ is the number of monomers on the chain, $\vec{r}_i$ is 
the position vector of the $i$th monomer, and $\vec{r}_{\rm cm}$ 
the center of mass of the chain.
The results are presented in Fig.~\ref{Cm_P1N48_Rg2.eps} where
the mean square radius of gyration $\langle R_{\rm g}^2 \rangle$ 
is plotted as a function of  salt concentration $C_{\rm s}$ 
at four monomer concentrations, $C_{\rm m}=0.00032$, $0.002$, $0.008$
and $0.016$.
The angular bracket $\langle \cdot \rangle$ means averaging over chain 
configurations.  Two behaviors were observed.
The first one is that each curve displays a similar trend of behavior
against salt concentration. 
Upon addition of salt, the chain undergoes two structural transitions:
a collapsed transition and a followed reexpanding transition,
as having been observed in the previous work~\cite{hsiao06a}.
The first transition results from the screening of the Coulomb
interaction by the added salt.
The second transition is still not well understood
although it is a result of the competition between energetic gain 
and entropic loose in counterion condensation and in shape 
transformation of polyelectrolytes
to minimize the free energy.
We observed that the minimum of $\langle R_{\rm g}^2 \rangle$ appears 
near the salt concentration $C_{\rm s}^{*}=C_{\rm m}/4$
and the more dilute the monomer concentration, the smaller value the minimum of 
$\langle R_{\rm g}^2 \rangle$ will be.
$C_{\rm s}^{*}$ is the equivalence point at which  
the amount of charges of the tetravalent counterions exactly
neutralizes the charges on polyelectrolytes
as if the salts and the polyelectrolytes are in the stoichiometric reaction.
We will call $C_{\rm s}^{*}$ the equivalence concentration.
In a salt-free solution ($C_{\rm s}=0$), we found that  
the chain size increases with decreasing monomer concentration
(refer to the four data points near the left vertical axis of the figure). 
This effect has been discussed in literatures~\cite{stevens95,liao03}  
and our results are in accordance with them.
And the second behavior is that $\langle R_{\rm g}^2 \rangle$ 
for different $C_{\rm m}$ attains approximately the same value 
while the salt concentration is larger than the equivalence concentration. 
Therefore,  the curves approximately overlap each other.
We noticed, furthermore, that in the region $C_{\rm s} \le C_{\rm s}^{*}$, 
the $\langle R_{\rm g}^2 \rangle$ curves for different $C_{\rm m}$ can also 
lie over each other under an appropriate transformation.
We demonstrate this overlap in the inset of Fig.~\ref{Cm_P1N48_Rg2.eps} 
where the abscissa is $C_{\rm s}/C_{\rm m}$ and 
the ordinate is $A_{\rm m}(\langle R_{\rm g}^2 \rangle-\langle R_{\rm g}^{*2} \rangle)$
with $\langle R_{\rm g}^{*2} \rangle$ the minimum value of $\langle R_{\rm g}^2 \rangle$.
$A_{\rm m}$ depends smoothly on $C_{\rm m}$ and, in the figure, 
is equal to $1$, $1.267$, $1.657$, and $2.049$ 
for $C_{\rm m}=0.00032$, $0.002$, $0.008$, and $0.016$, respectively.  

The size of polyelectrolytes in multivalent salt solutions has been studied by 
small-angle scattering experiments~\cite{dubois01}.
The results showed that it diminishes with increasing or decreasing $C_{\rm s}$ 
toward the region where the condensation of polyelectrolytes occurs
(cf.~Table 2 of Ref.~\cite{dubois01}). 
The results of our simulations are in accordance with the experiments.
We make a hypothesis that chain size can be served as a qualitative condition 
to distinguish between a monophasic solution of polyelectrolyte and a diphasic one. 
More precisely, if the chain size is larger than some value, the chains are
dissolved in the solution and the system is in a homogeneous phase;
if it is smaller, the chains are precipitated from the solution and
a phase separation occurs.
Under this hypothesis,  polyelectrolyte solutions of different $C_{\rm m}$ become 
monophasic as $C_s$ is increased over a concentration in the high-salt region,
since the chain size exceeds the some value nearly at the same $C_{\rm s}$
due to the similarity of the $\langle R_{\rm g}^2 \rangle$ curves 
against $C_{\rm s}$ in this region.
Therefore, the upper boundary of the salt concentration window 
for the condensation of polyelectrolyte is independent of $C_{\rm m}$.
Similarly, the lower boundary of the window approximately depends linearly on 
$C_{\rm m}$, owing to the similarity of the 
$A_{\rm m}(\langle R_{\rm g}^2 \rangle-\langle R_{\rm g}^{*2} \rangle)$
curves against $C_{\rm s}/C_{\rm m}$ in the low-salt region;
the weak dependence of $A_{\rm m}$ and $\langle R_{\rm g}^{*2} \rangle$ on
$C_{\rm m}$ ensures that the chain size surpasses the some value
roughly at the same $C_{\rm s}/C_{\rm m}$. 
Therefore, Fig.~\ref{Cm_P1N48_Rg2.eps} provides an essential information 
to understand the boundaries of reentrant condensation.   
  
The aim of this article is to understand the properties of multiple flexible 
polyelectrolytes with added salt in an aqueous solution.
It is known that the Bjerrum length $\ell_{\rm B}$ in water 
is equal to 7.14\AA\@ at room temperature.  
Our simulation setup, hence, corresponds to a collision diameter $\sigma_{\rm m}$ of 
monomer equal to 2.38\AA\@.  
This model represents a polyelectrolyte system 
with linear charge density equal to a unit charge per 
$2.62$\AA\@\footnote{The average bond length in our simulations is about 
1.1 $\sigma_{\rm m}$.}, and can be employed to simulate a strongly-charged system 
such as sodium polystyrene sulfonate. 
The four monomer concentrations studied in this section,
while converted to real unit, are equal to 0.039M, 0.246M, 0.986M, and 
1.971M, respectively. 
Even the largest $C_{\rm m}$ is still smaller than 
the overlap threshold $C_{\rm m}^{*}$ estimated by $3 N/(4\pi R_{\rm g}^3)$. 
Therefore, our systems are dilute polymer solutions. 
Readers should be aware that the value up to nearly 2M is a very high monomer 
concentration for a real polymer system, and our systems are \textit{dilute} only 
in the formal meaning of the world since the chains are very short.
In the following sections, we fix $C_{\rm m}$ at 0.008 and  
simulate a system comprising multiple chains.
This monomer concentration is one or two order of magnitude
higher than a typical monomer concentration used in 
experiments~\cite{delacruz95,raspaud98,dubois01}.
However, based on the fact that systems at different monomer concentrations
reveal similar evolution of chain size against salt concentration, 
the results obtained at $C_{\rm m}=0.008$ can be used 
to understand a system at more dilute $C_{\rm m}$.
We mention that this $C_{\rm m}$ is a  choice so that the 
simulation box is not so big.
Therefore, the amount of the added salt is controllable under limited 
computing resources. 
Consequently, to study the behavior of polyelectrolytes in
salt solutions, covering a broad range of salt concentration, 
becomes numerically feasible.

\subsection{Shape of polyelectrolytes}
\label{Sec_Shape}
In this section and in the followings, the studied system contains four polyelectrolytes 
and the monomer concentration is fixed at $C_{\rm m}=0.008$. 
This section is devoted to the study of chain conformation 
in tetravalent salt solutions.
The effects of salt concentration and size of tetravalent counterions 
are investigated.
We vary the size of tetravalent counterion, 
$\sigma_{+4}$, from 0.0 to 4.0 and 
fix the size of the other ions at 1.0.
A large $\sigma_{+4}$ can represent a large cation, 
a large ion group, or a large charged colloid. 
It can be also a result of the hydration of a small ion. 
A small $\sigma_{+4}$ may be less representative for real systems 
but theorists are generally interested in it because many models 
are established on the base of vanishing ion size.

\begin{figure}[htbp] 
\begin{center}
\includegraphics[width=0.4\textwidth,angle=270]{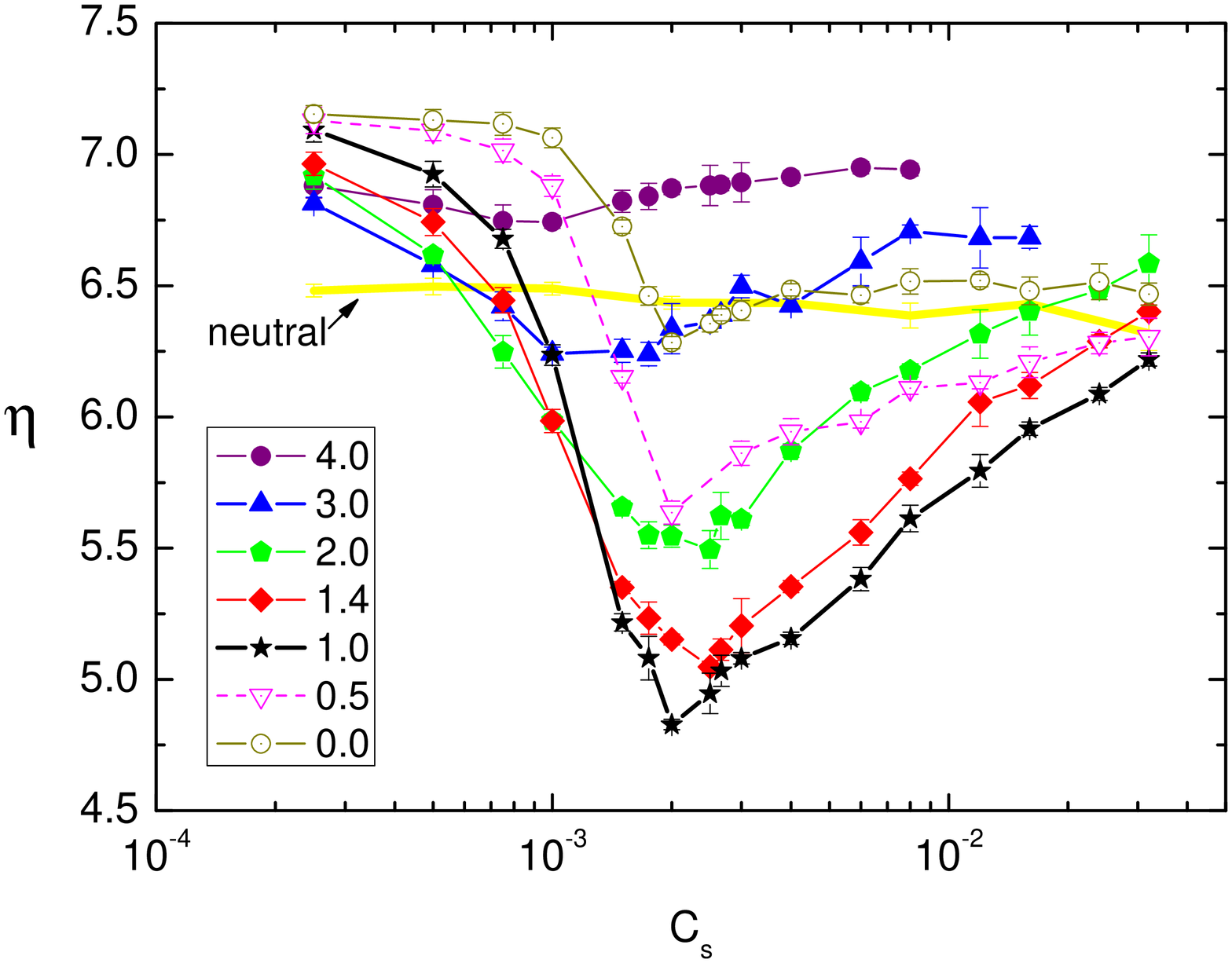}
(a)
\includegraphics[width=0.4\textwidth,angle=270]{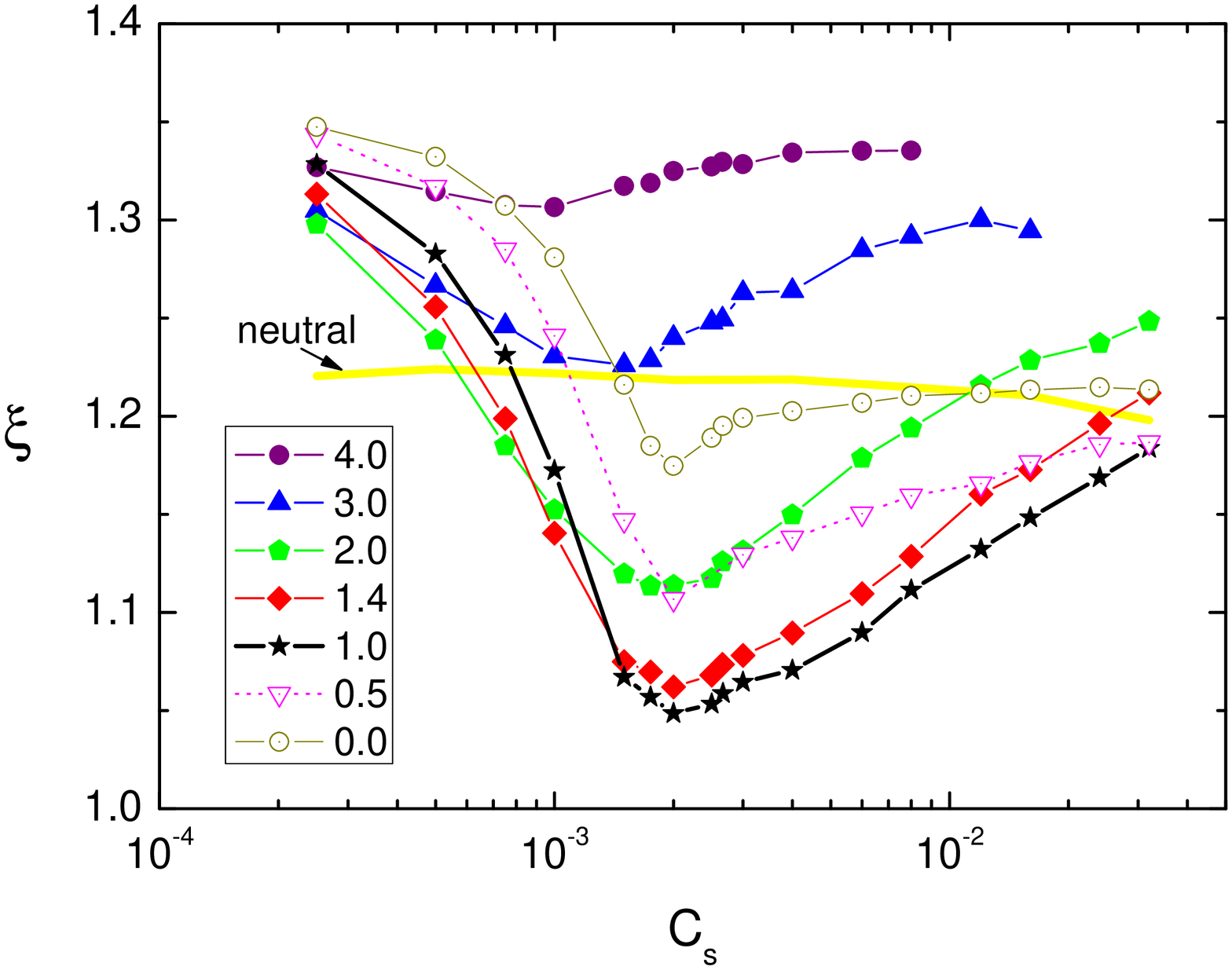}
(b)
\caption{(a) $\eta\equiv\langle R_{\rm e}^2\rangle/\langle R_{\rm g}^2\rangle$ as a function of $C_{\rm s}$,  and
(b) $\xi\equiv\langle R_{\rm g}^2\rangle^{1/2}/\langle R_{\rm h}\rangle$ as a function of $C_{\rm s}$, 
for the cases of different size $\sigma_{+4}$ of the tetravalent counterions. 
The symbol used for each $\sigma_{+4}$ is
indicated at the left-bottom side of the figures. The error bars are smaller than
the size of the symbols.  The result for neutral
polymers is depicted as reference and marked by  ``neutral''.  }
\label{shapeP4N48_eta1&xi1.eps}
\end{center}
\end{figure}

Morphology of a polymer can be quantified by many quantities.
The first quantity that we studied is the shape factor $\eta$, 
defined as the ratio of the mean-square end-to-end distance, 
$\langle R_{\rm e}^2 \rangle=\langle (\vec{r}_1-\vec{r}_N)^2 \rangle$,
to the mean-square radius of gyration $\langle R_{\rm g}^2 \rangle$.  
$\eta$ attains a value 12 for a rodlike polymer, 6.3 for a flexible
chain in a good solvent, and 6 for an ideal chain~\cite{stevens95}. 
For a compact structure, the value of $\eta$ is small.
For example, if the chain has a spherical morphology 
and the two ends are randomly positioned, $\eta$ is equal to 2. 
In Fig.\ref{shapeP4N48_eta1&xi1.eps}(a), we show how the shape factor
$\eta$ varies with $C_{\rm s}$ at different $\sigma_{+4}$.
In the figure, a reference curve, marked by ``neutral'',  is 
depicted. 
It represents a system composed of four neutral polymers of 
48 monomers in tetravalent salt ($\sigma_{+4}=1.0$) solutions.
Focus firstly on the polyelectrolyte solutions with $\sigma_{+4} =1.0$, 1.4 and 2.0.
We witnessed that $\eta$ decreases at first and then increases upon addition of salt,
in analogy of the radius of gyration studied in the previous section,
and shows a V-shaped curve against $C_{\rm s}$ in the semilog plot.
In the low-salt region or in the high-salt region, $\eta$ is larger than or
close to the value 6.3, indicating that the chains have elongated or coil-like 
structures.
In between, it appears a minimum near the equivalence 
concentration $C_{\rm s}^{*}=0.002$. 
The value of $\eta$ tells us that the chains possess compact 
structures but not so compact as spheres.
For the cases with larger $\sigma_{+4}$ ($\sigma_{+4}=3.0$ and $4.0$), 
$\eta$ shows a shallow minimum and the whole curve stays basically above 
the value 6.3. In these cases, the chains are in an extended state.
It is worth to notice that for $\sigma_{+4} \ge 1.0$, 
$\eta$, in principle, increases with $\sigma_{+4}$ at a given $C_{\rm s}$. 
Now pay attention to the cases with $\sigma_{+4}<1.0$.
$\eta$ displays a similar V-shaped curve against $C_{\rm s}$  
but opposite to the previous cases,
its value increases with decreasing $\sigma_{+4}$.
Moreover, the degree of chain reexpansion in the high-salt region 
is weakened while $\sigma_{+4}$ is decreased.
Noticeably, for the case $\sigma_{+4}=0.0$, 
$\eta$ displays a sharp transition from a value 7.2 in the low-salt solution to 
a value 6.5 in the vicinity of $C_{\rm s}^{*}$.
In the latter region, the chains behave similarly to neutral polymers.

The second quantity that we studied is $\xi$, which is the ratio of 
the radius of gyration $\langle R_{\rm g}^2 \rangle^{1/2}$ to 
the hydrodynamic radius $\langle R_{\rm h}\rangle$
computed by $R_{\rm h}^{-1}=N^{-2}(\sum_{i=1}^{N}\sum_{j=1,j\neq i}^{N} r_{ij}^{-1})$.
$\xi$ is equal to $\sqrt{3/5}\simeq 0.775$ for a hard sphere and
attains a value 2.25 for a rodlike chain~\cite{ou05}.
Renormalization group theory has predicted $\xi \simeq 1.56$ 
in a good solvent and $\xi \simeq 1.24$ at $\Theta$ point~\cite{oono83}, 
both of which have been shown in agreement 
with experiments~\cite{rubinstein03book}. 
We present in Fig.~\ref{shapeP4N48_eta1&xi1.eps}(b) 
the variation of $\xi$ against $C_{\rm s}$.
The referenced curve for the system of neutral polymers is depicted.
The curve was found very close to the value 1.24, 
which seems to indicate an ideal-chain behavior.  
However, the neutral chains should behave as random coils and 
the value of $\xi$ would be larger than 1.24. 
The deviation of $\xi$ from the value of a coil is probably 
due to the short chain length used in our simulations.  
D{\"u}nweg \textit{et al.} have studied the finite size effect and 
shown that the dependence of $\xi$ on chain length is not 
negligible~\cite{dunweg02}.
Therefore, $\xi$ reported here departs slightly from the asymptotic value 
of an infinite chain length. 
Nonetheless, the results are still indicative and useful  
to understand how $\xi$ varies with salt concentration.
In the small $C_{\rm s}$ region, the polyelectrolytes are more extended 
than the neutral polymers.
In the middle region of $C_{\rm s}$, the chains attain compact structures 
when $0.5\le\sigma_{+4}\le 2.0$.  
The dependence of $\xi$ upon $\sigma_{+4}$ resembles to that of $\eta$. 
For large $\sigma_{+4}$ ($\sigma_{+4}>2.0)$, the chains are elongated.
For vanishing $\sigma_{+4}$, we observed that following a sharp decrease, 
$\xi$ reattains approximately a constant value, close to that for 
the neutral polymers.

\begin{figure}[htbp] 
\begin{center}
\includegraphics[width=0.4\textwidth,angle=270]{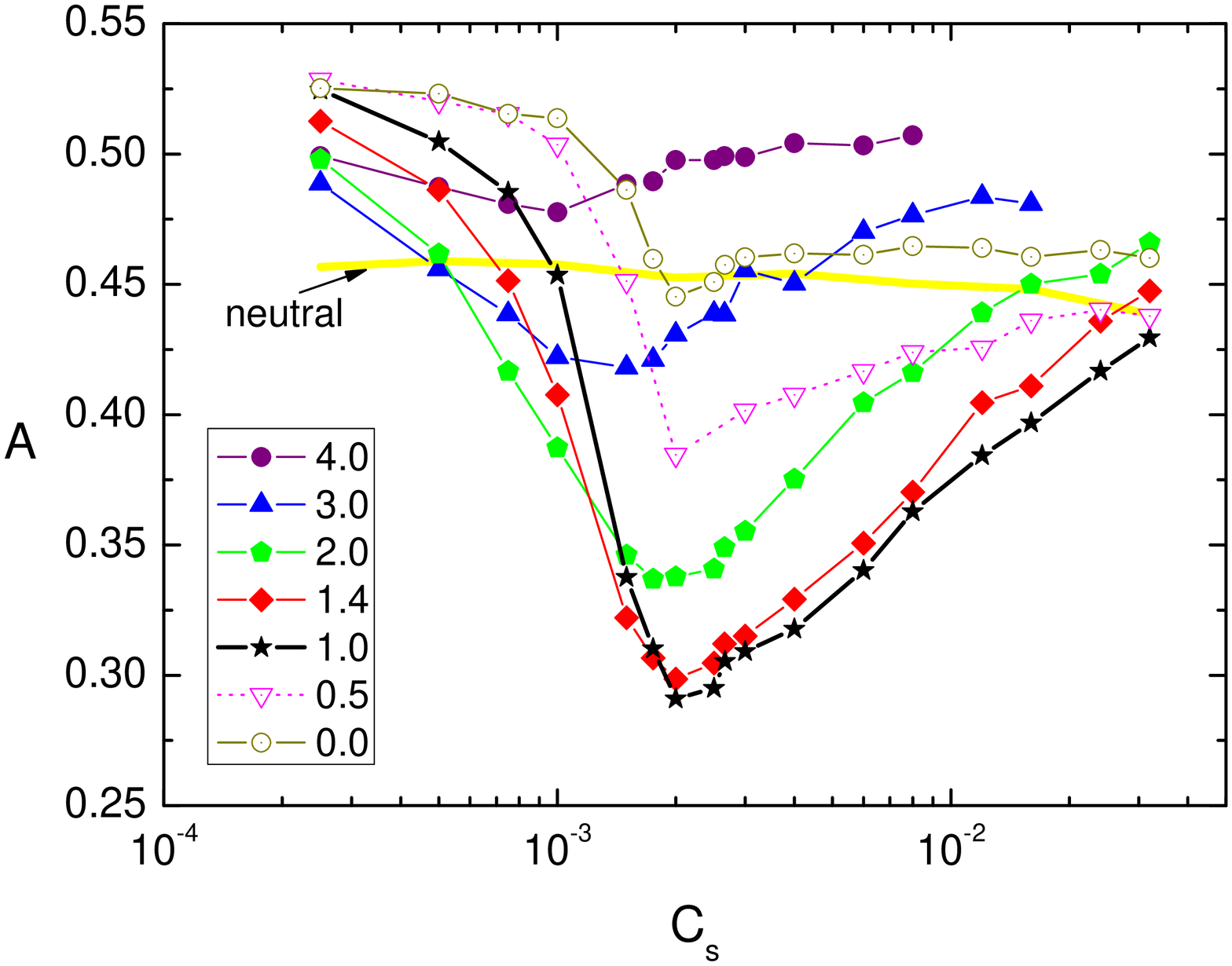}
(a)
\includegraphics[width=0.4\textwidth,angle=270]{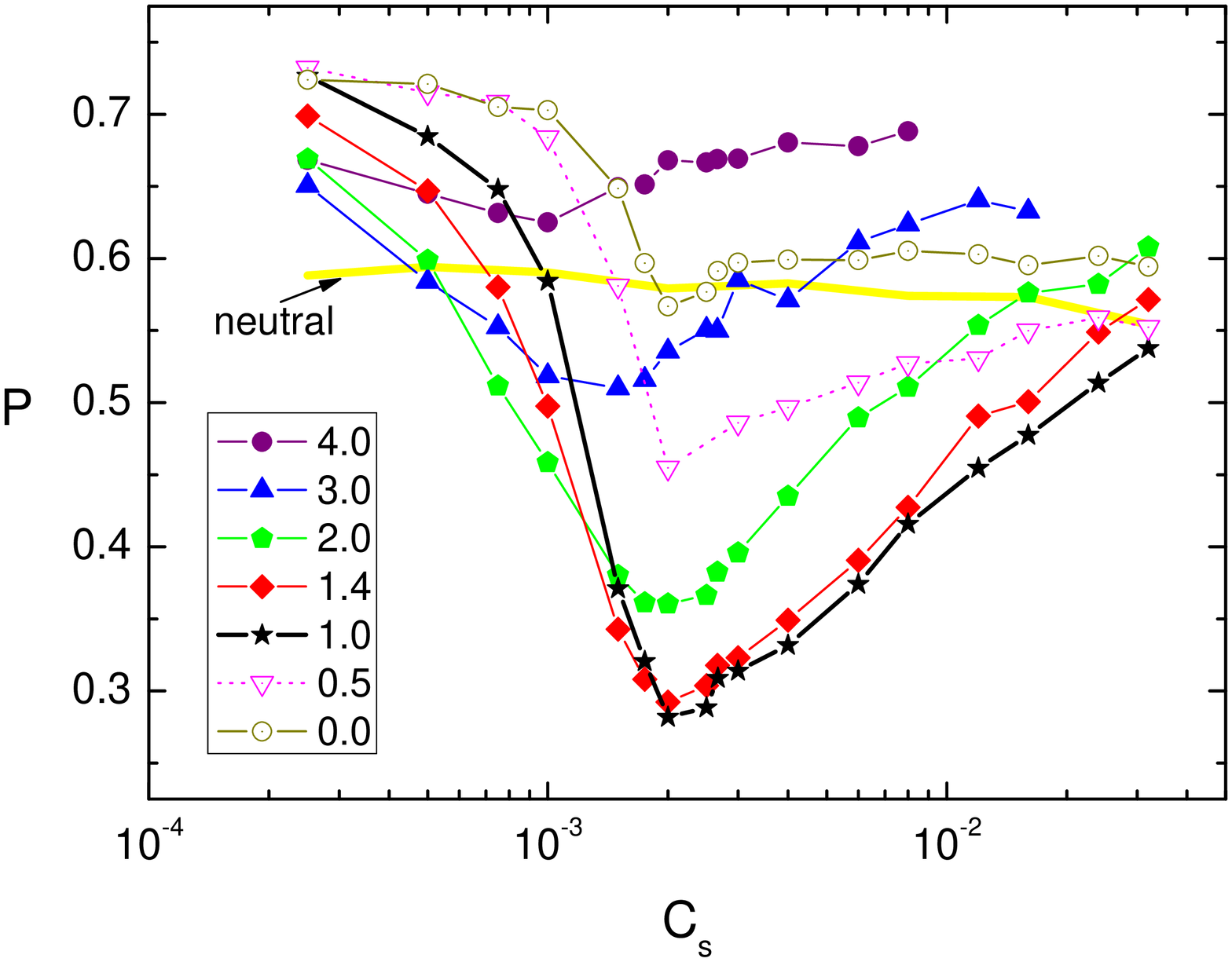}
(b)
\caption{(a) $A$ as a function of $C_{\rm s}$,  and
(b) $P$ as a function of $C_{\rm s}$, for different $\sigma_{+4}$. 
The symbol used for each $\sigma_{+4}$ is indicated at the left-bottom 
side of the figures. The error bars are smaller than
the size of the symbols.  The result for neutral
polymers is depicted as reference and marked by  ``neutral''.  }
\label{shapeP4N48_Asph&Sp.eps}
\end{center}
\end{figure}

A more fundamental way to study the shape of a polymer is to investigate the tensor
of radius of gyration ${\cal{T}}$~\cite{solc71}, 
defined by: 
\begin{equation}
{\cal{T}}_{\alpha\beta}=\frac{1}{2N^2}\sum_{i=1}^{N}\sum_{j=1}^{N}
(r_{i\alpha}-r_{j\alpha})(r_{i\beta}-r_{j\beta})
\end{equation}
where $\alpha$, $\beta=1,2,3$ denote the three Cartesian components and
$r_{i\alpha}$ is the $\alpha$ component of the position of the $i$th monomer.
Let $\lambda_1$, $\lambda_2$ and $\lambda_3$ be the three eigenvalues of
${\cal T}$.
A quantity called ``asphericity''~\cite{christos89}, 
which measures the deformation from a spherical geometry, is defined by
\begin{equation}
A=\frac{1}{2}\left\langle 
\frac{(\lambda_1-\lambda_2)^2+(\lambda_2-\lambda_3)^2+(\lambda_3-\lambda_1)^2}
{(\lambda_1+\lambda_2+\lambda_3)^2}\right\rangle.
\end{equation}
It takes a value between 0 (sphere) and 1 (rod). 
For a coil polymer, it is 0.431 obtained from simulations~\cite{bishop88}. 
Another quantity describing the degree of prolateness is defined by
\begin{equation}
P=\left\langle \frac{(\lambda_1-\bar{\lambda})(\lambda_2-\bar{\lambda})(\lambda_3-\bar{\lambda})}
{\bar{\lambda}^3}\right\rangle
\end{equation}
where $\bar{\lambda}=(\lambda_1+\lambda_2+\lambda_3)/3$~\cite{christos89}.
It is ranged from $-1/4$ to $2$ and can be used to distinguish between a prolate shape ($0<P<2$) 
and an oblate one ($-\frac{1}{4}<P<0$).
For neutral polymers in the dilute limit, simulations predict a prolate chain with
$P=0.541$~\cite{jagodzinski92} .
We present in Figs.~\ref{shapeP4N48_Asph&Sp.eps}(a) and \ref{shapeP4N48_Asph&Sp.eps}(b) 
the asphericity $A$ and the prolateness parameter $P$ as a function of $C_{\rm s}$. 
Resembling to the behaviors of $\eta$ and $\xi$, 
$A$ and $P$ show a minimum value in the vicinity of $C_{\rm s}^{*}$. 
The curve associated to $\sigma_{+4}=1.0$ roughly separates two different behaviors 
of evolution: 
$A$ and $P$ increase with increasing and with decreasing $\sigma_{+4}$ from 1.0.
The value of $A$ lies between 0.275 and 0.55. 
Therefore, even in a compact structure (where $C_{\rm s}$ is near $C_{\rm s}^{*}$),
the chain shape is asymmetric.
In a low-salt solution, the chains show extended 
morphology but are not as elongated as rods. 
On the other hand, $P$ is larger than zero for all of the studied cases, 
indicating that the polyelectrolytes display prolate conformation.  
At a high salt concentration, $A$ is close to 0.45 and $P$ is close to 0.54, 
suggesting once more that the polyelectrolytes behave as the neutral polymers 
in the high-salt region.  
For the case with null $\sigma_{+4}$, chain reexpansion does not occur;
$A$ and $P$ stay at a constant value when $C_{\rm s}$ goes beyond $C_{\rm s}^{*}$.   
The results obtained in this section strongly suggest that $\sigma_{+4}=1.0$ is 
the optimal condition to pack polyelectrolytes into the smallest volume.

\subsection{Single-chain structure factor and swelling exponent}
\label{Sec_SF}
\begin{figure}[htbp] 
\begin{center}
\includegraphics[width=0.4\textwidth,angle=270]{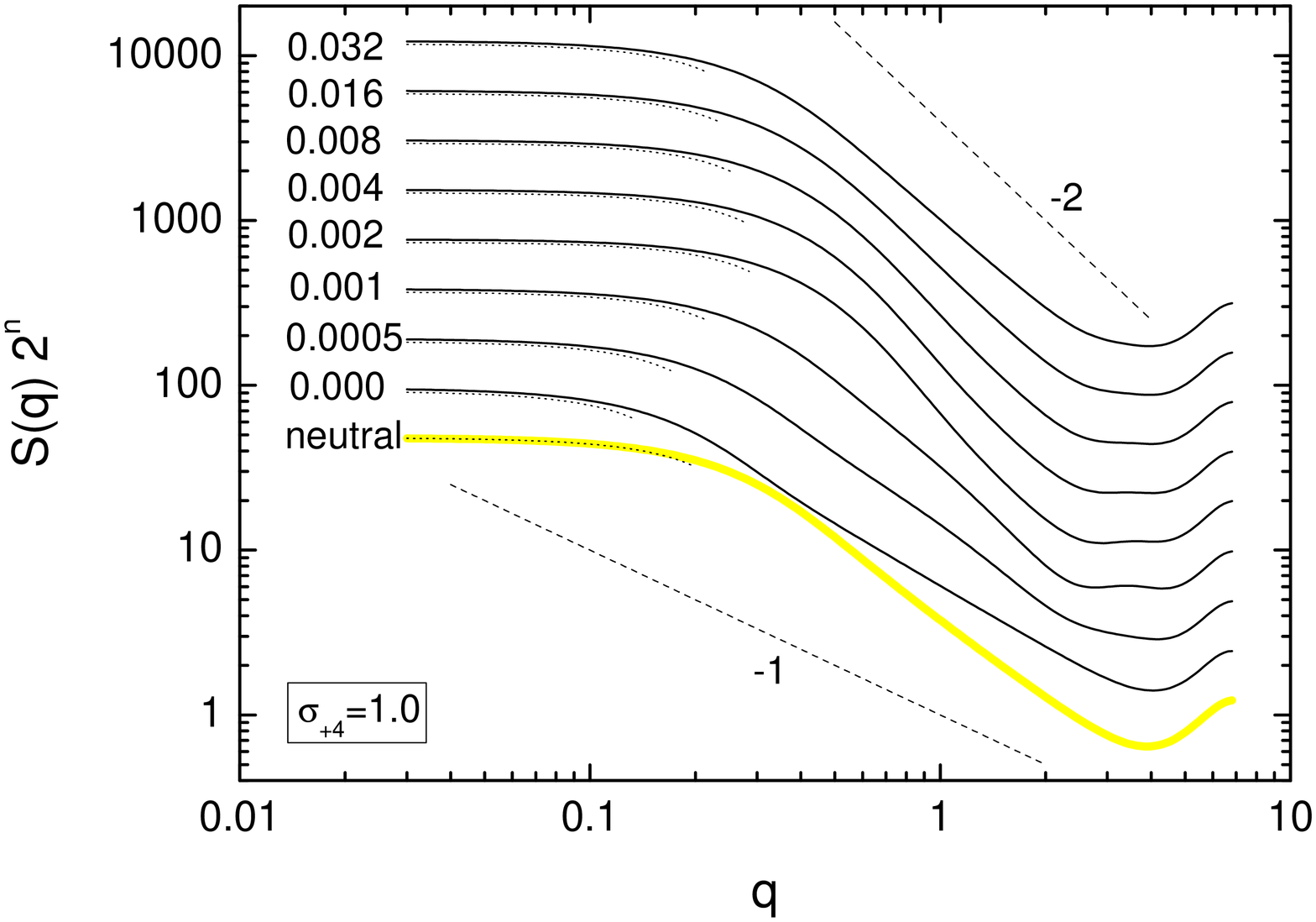}
\\
(a)
\\
\includegraphics[width=0.4\textwidth,angle=270]{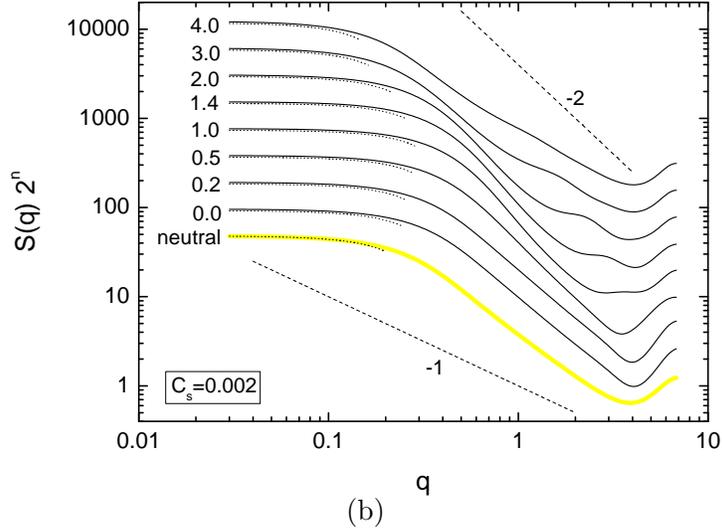}
\\
(b)
\\
\caption{(a) $S(q)$ for $\sigma_{+4}=1.0$ at different $C_{\rm s}$
and (b) $S(q)$ at $C_{\rm s}=0.002$  for different  $\sigma_{+4}$.
The value of $C_{\rm s}$ or $\sigma_{+4}$, at which the simulations were
running, is indicated at the left side of each corresponding curve.
$S(q)$ for the neutral polymers at $C_{\rm s}=0.002$ is depicted as 
reference and marked by ``neutral''.
For clarity, the curves, from bottom to top, have been multiplied by a factor
$2^n$ with $n$ started from zero and increased by one for every curve.
The dotted curve, near each $S(q)$ and terminated at $q=R_{\rm g}^{-1}$, 
is the Guinier function $N(1-q^2\langle R_{\rm g}^2\rangle/3)$.
Two dashed lines, which represent two power-law dependences, $q^{-1}$ and $q^{-2}$
(marked by ``-1'' and ``-2'', respectively), are drawn in the plots.}
\label{SFP4N48_Sxxx&txxx}
\end{center}
\end{figure}

In this section, we investigate the single-chain structure factor.
This quantity is experimentally accessible and provides an essential information 
to theoretical analysis.  The single-chain structure factor is defined by 
\begin{equation}
S(\vec{q})=
\frac{1}{N_{\rm p}N} \sum_{p=1}^{N_{\rm p}}
\left< \left|\sum_{j=1}^{N} \exp(i\vec{q}\cdot\vec{r}_{p,j})\right|^2 \right>
\end{equation}
where $\vec{q}$ is the scattering wavevector, $N_{\rm p}=4$ is the number of 
polymer chains, $N=48$ is the number of monomers on a chain, and  
$\vec{r}_{p,j}$ is the position of the $j$th monomer on the $p$th chain. 
Suppose that the system has spherical symmetry. 
The dependence of the structure factor on the orientation of $\vec{q}$ can be 
therefore integrated out: $S(q)=\frac{1}{4\pi}\int d\Omega\, S(\vec{q})$ where 
$q$ is the norm of $\vec{q}$ and $\Omega$ is the solid angle.
This integration was performed numerically by sampling over a set of 
randomly-oriented $\vec{q}$ vectors.
We present, in Fig.~\ref{SFP4N48_Sxxx&txxx}(a), $S(q)$ at several salt concentrations 
for the case of $\sigma_{+4}=1.0$.
For the reason of clarity, the curves in the figure have been multiplied by $2^{n}$ 
with $n$ being increased by one for each curve from the bottom to the top. 
$S(q)$ for the neutral polymers is depicted as reference with $n=0$.
We witnessed that $S(q)$ shows typical behavior.
In the Guinier regime ($qR_{\rm g}\ll 1$), 
$S(q)$ behaves similarly to $N(1-q^2\langle R_{\rm g}^2\rangle/3)$.
It is demonstrated by plotting $N(1-q^2\langle R_{\rm g}^2\rangle/3)$ 
(in dotted curve) near the corresponding $S(q)$ with the value of 
$\langle R_{\rm g}^2\rangle$ adopted from Sec.~\ref{Sec_Shape}.
The dotted curve terminates at $q=R_{\rm g}^{-1}$ in order to indicate 
the boundary of the Guinier regime. 
In the regime $R_{\rm g}^{-1} \ll q \ll \sigma_{\rm m}^{-1}$, 
$S(q)$ shows power-law-like behavior, manifested by 
a linear dependence of $\log S(q)$ on $\log q$.
The slope $s$ of the linear dependence is related to the swelling exponent 
$\nu$ by the relation $\nu=-1/s$, 
which describes the size extension of a polymer: $R_{\rm g} \sim N^{\nu}$. 
In the regime $q>\sigma_{\rm m}^{-1}$, $S(q)$ displays non-universal behavior. 
However, while $q \gg \sigma_{\rm m}^{-1}$, self-scattering of monomer is the
only contribution, leading, therefore, $S(q)$ to a universal value $1$.
We mention that the Krathy regime, in which $S(q) \sim q^{-2}$, is not observed 
in our study because the system is a dilute solution and 
the chain length is not long enough~\cite{muller00}.
For $\sigma_{+4}$ other than 1.0, we observed analogous trends of behavior.
For example, the variation of $S(q)$ for different $\sigma_{+4}$ 
at the equivalence concentration is presented in Fig.~\ref{SFP4N48_Sxxx&txxx}(b), in which  
we rediscovered the three regimes: the Guinier regime, the power-law-like regime
and the non-universal regime.
We noticed that in the last regime, a small bounce appears on the curve and,
with increasing the size of tetravalent ions, gradually moves toward small $q$.

\begin{figure}[htbp] 
\begin{center}
\includegraphics[width=0.4\textwidth,angle=270]{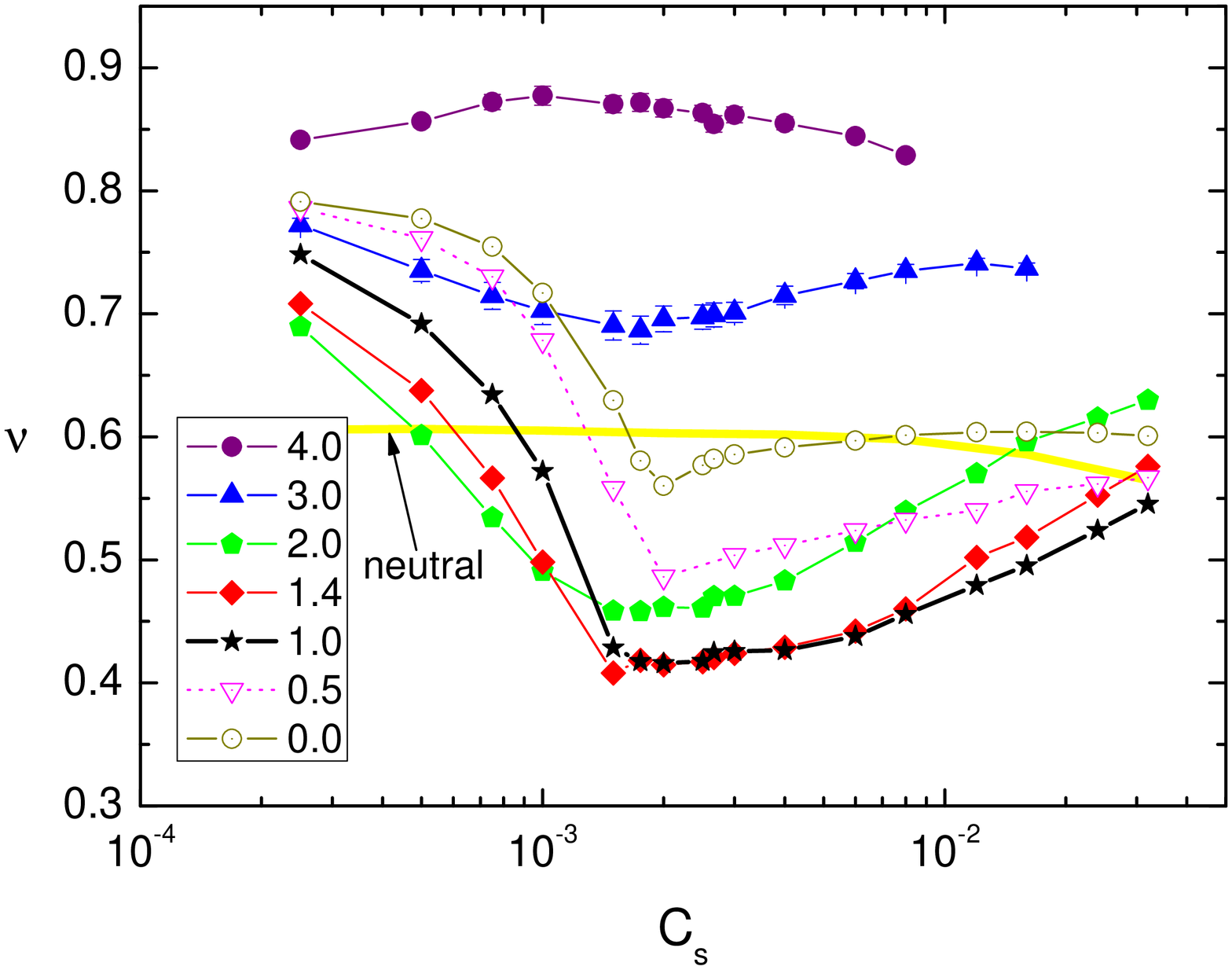}
\caption{$\nu$ as a function of $C_{\rm s}$ for different $\sigma_{+4}$.
The symbol used for each $\sigma_{+4}$ is indicated at the left-bottom side of the figure.
$\nu$ for the referenced neutral-polymer system is depicted and marked by ``neutral''.  }
\label{nuP4N48.eps}
\end{center}
\end{figure}

The swelling exponent $\nu$ of the polyelectrolytes was calculated by
performing least-square fits in the log-log plot of $S(q)$ in the power-law-like regime.
The fitting region was chosen to be $0.63<q<1$ 
and the fitting reliability has been verified to be excellent.
We present the results of the swelling exponent $\nu$   
in Fig.~\ref{nuP4N48.eps}. 
$\nu$ for the referenced neutral polymers was verified at first and
found to attain a value close to the Flory's one, $0.6$. 
Next, we focused on the swelling exponent for the polyelectrolytes.
We observed that $\nu$ exceeds the value for the neutral polymers at 
a low salt concentration. 
It indicates a strong extension of chain size due to Coulomb repulsion 
between monomers.
For the cases with $\sigma_{+4}=0.5$, 1,0, 1,4, and 2.0,
the swelling exponent shows a V-shaped curve against $C_{\rm s}$: 
$\nu$ decreases and then increases with increasing salt concentration.
A minimum occurs near the equivalence concentration $C_{\rm s}^*$
and, at this moment, the chains show compact structures but are less compact 
than spheres because a spherical structure scales as $N^{1/3}$.
We remark that the swelling exponent obtained here for $\sigma_{+4}=1.0$
is consistent with that obtained by direct variation of the chain length 
of a system containing a single polyelectrolyte~\cite{hsiao06b}
except in the mid-salt region where $\nu$ in the latter study is 
slightly smaller and acquires a value $1/3$ at $C_{\rm s}^*$.
This deviation is due to multi-chain aggregation occurred
in our system, which deforms the chain shape from a sphere 
and, consequently, increases $\nu$. 
For the case with vanishing $\sigma_{+4}$, 
following a drastic decrease, $\nu$ retains roughly a value 0.6
in the high-salt region and the chains, hence, scale 
like the neutral polymers.
For the cases with large $\sigma_{+4}$ ($\sigma_{+4}=3.0$ and 4.0),
$\nu$ is greater than 0.6 but still less than 1.0. 
The chains, hence, do not reveal rigid rod-like structures.
In principle, $\sigma_{+4}=1.0$ separates two behaviors of evolution 
at a salt concentration: $\nu$ increases with increasing and with  
decreasing $\sigma_{+4}$ from 1.0. 
The results demonstrate again that the polyelectrolytes
are mostly condensed while the size of the tetravalent counterions
is compatible with the one of the monomers.

\subsection{Persistence length}
\label{Sec_lp}
\begin{figure}[htbp] 
\begin{center}
\includegraphics[width=0.4\textwidth,angle=270]{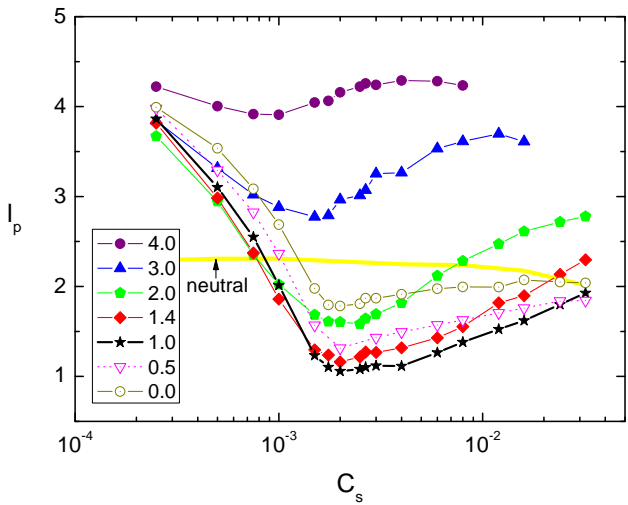}
\caption{$\ell_{\rm p}$ as a function of $C_{\rm s}$ for different  
$\sigma_{+4}$. The symbol used for each $\sigma_{+4}$ is
indicated at the left-bottom side of the figure. The error bars are smaller than 
the size of the symbols.  The bare persistence length 
$\ell_{0}$ is  depicted as reference and marked by  ``neutral''. }
\label{lpP4N48.eps}
\end{center}
\end{figure}

There is a considerable interest to understand the variation of
the persistence length of a polyelectrolyte against salt concentration.
Persistence length is a local property, which measures the 
stiffness of a polymer, and is distinguished from global conformational quantities.  
There are many ways to define the persistence length~\cite{ullner97,ullner02}. 
In this section, we adopt a microscopic definition to calculate it:
\begin{equation}
\ell_{\rm p}= \frac{1}{2b} \sum_{i=0}^{(N/2)-1} 
\langle \vec{b}_{N/2}\cdot\vec{b}_{(N/2)-i}
+\vec{b}_{N/2}\cdot\vec{b}_{(N/2)+i} \rangle
\label{persist_lp}
\end{equation}
where $\vec{b}_{j}=\vec{r}_{j+1}-\vec{r}_{j}$ is the $j$th 
bond vector and $b=\langle \vec{b}_j^2 \rangle^{1/2}$ is the root-mean-square
bond length.
In this definition, the contribution of the finite bond length, 
which corresponds to the $i=0$ term in Eq.~\ref{persist_lp}, is considered. 
The obtaining $\ell_{\rm p}$  as a function of $C_{\rm s}$ 
for different $\sigma_{+4}$ is plotted in Fig.~\ref{lpP4N48.eps}. 
Resembling to the quantities of chain conformation, $\ell_{\rm p}$ decreases
in the chain collapse region and increases in the chain reexpansion region. 
A minimum value appears in the vicinity of $C_{\rm s}^{*}$.
The degree of increase diminishes with decreasing $\sigma_{+4}$. 
For vanishing  $\sigma_{+4}$,  it is so weak that 
$\ell_{\rm p}$ is roughly a constant as $C_{\rm s}>C_{\rm s}^*$. 
Similar to the results in the previous sections, $\ell_{\rm p}$ increases 
with both increasing and decreasing $\sigma_{+4}$  from 1.0.  
$\sigma_{+4}=1.0$, therefore, denotes a condition to separate the two behaviors of evolution 
of $\ell_{\rm p}$.

The persistence length has generally two contributions: 
the bare persistence length $\ell_{0}$ (in the absence of electrostatic force)
and the electrostatic persistence length $\ell_{\rm e}$ (owing to 
the Coulomb interaction) \cite{odijk77,skolnick77}. 
In this study, the bare persistence length $\ell_{0}$ was calculated 
by performing simulations on the referenced system 
where four neutral chains and 192 neutral particles
are placed in salt solutions. 
The result is depicted in Fig.~\ref{lpP4N48.eps} and the curve is marked by ``neutral''.
We found that $\ell_{0}$ is not sensitive to the salt concentration 
and roughly attains a value 2.3.
It slightly decreases in the high-salt region, probably due to the jamming effect 
by the salt ions. 
For the polyelectrolyte solutions with large $\sigma_{+4}$ ($\sigma_{+4}=3.0$ and 4.0), 
the $\ell_{\rm p}$ curve thoroughly lies above the $\ell_{0}$ curve, 
indicating a more rigid flexibility than a neutral polymer. 
The electrostatic persistence length is, hence, positive. 
On the other hand, for the polyelectrolyte solutions with $\sigma_{+4}\le 2.0$, 
$\ell_{\rm p}$ is smaller than $\ell_{0}$ in the mid-salt region and 
shows one or two intersections with the $\ell_{0}$ curve.
Therefore, $\ell_{\rm e}=\ell_{\rm p}-\ell_{0}$ is positive in 
a salt-free solution and becomes eventually negative upon addition of salt.
It increases in the high-salt region and reattains a positive value 
for $\sigma_{+4}=1.4$ or $2.0$.

\begin{figure}[htbp] 
\begin{center}
\includegraphics[width=0.4\textwidth,angle=270]{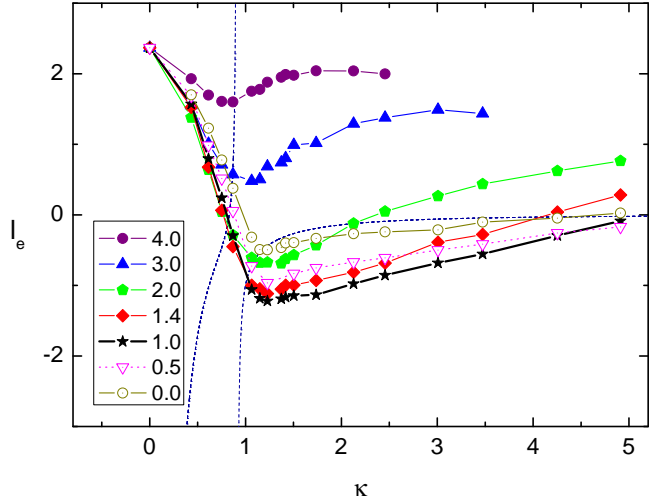}
\caption{$\ell_{\rm e}$ as a function of $\kappa$ for different $\sigma_{+4}$. 
The symbol used for each $\sigma_{+4}$ is indicated at the left-bottom 
side of the figure. The error bars are smaller than 
the size of the symbols.  The dashed curve depicts $\ell_{e}$ 
predicted by Eq.~(\ref{le_ariel}). }
\label{leP4N48.eps}
\end{center}
\end{figure}

The negative contribution of electrostatics to the persistence length 
has been experimentally observed~\cite{baumann97,hugel01}.
Recently Ariel and Andelman (AA) pointed out that a negative $\ell_{\rm e}$ 
indicates a mechanical instability (collapse) of polyelectrolytes
due to the presence of multivalent counterions~\cite{ariel03b}. 
By taking into account the thermal fluctuations and correlations between 
bound counterions,  they derived an expression of $\ell_{\rm e}$
for an intrinsically-rigid charged polymer in ($Z$:1)-salt solutions 
in the limit of $b \ll \kappa^{-1} \ll (N-1)b$:
\begin{equation}
\ell_{\rm e}=
\left( \Gamma ( 2-\Gamma)-\frac{(\Gamma -1)^2}{\Gamma \ln(\kappa b)} \right)
\ell_{\rm e}^{\rm OSF} \,
\label{le_ariel}
\end{equation} 
where
$\Gamma=Z\ell_{\rm B}/b$ is the Coulomb strength parameter between
a monomer and a $Z$-valent counterion,  
$\kappa=\sqrt{4\pi Z(Z+1)\ell_{\rm B} C_{\rm s}}$ is the inverse 
Debye-H\"uckel screening length,
and $\ell_{\rm e}^{\rm OSF}=(4Z^2\kappa^2\ell_{\rm B})^{-1}$ is the 
electrostatic persistence length deduced by 
Odijk-Skolnick-Fixman (OSF) theory~\cite{odijk77,skolnick77}.
In the derivation, they have assumed point-like charges for the ions.
We present, in Fig.~\ref{leP4N48.eps}, $\ell_{\rm e}$ obtained from 
the simulations and that predicted by Eq.~(\ref{le_ariel}) 
as a function of $\kappa$. 
We witnessed that $\ell_{\rm e}$ depends strongly on the size of the tetravalent ions.
The negative regime of $\ell_{\rm e}$ has been successfully captured 
in the simulations.
To make a comparison with the theoretical prediction, 
we focused on the case with vanishing $\sigma_{+4}$. 
Please notice that this case does not exactly fulfill the assumption of the theory 
because in the simulations, the monovalent counterions and coions are not 
point charge-like and, also, the chains are not rigid.
However, according to our experiences~\cite{hsiao06a},
ions other than multivalent counterions do not play a decisive 
role on the properties of polyelectrolytes. 
This comparison, therefore, makes sense 
if the effect of chain rigidness is negligible.
We emphasize that the electrostatic correlations are fully taken into account 
in the simulations. 
In our study, the region in which Eq.~(\ref{le_ariel}) is valid is 
$0.02 \ll \kappa \ll 0.91$ because $Z=4$, $\ell_{\rm B}=3$, 
and $b$ is about 1.1.
Notice that Eq.~(\ref{le_ariel}) has a singularity at 
$\kappa=1/b \simeq 0.91$ and, hence, the curve is divided into two branches.
The left branch is located in the valid region but
fails to predict $\ell_{\rm e}$. 
It gives not only a wrong value but also a wrong trend of 
variation of $\ell_{\rm e}$ against $\kappa$. 
We noticed that Eq.~(\ref{le_ariel}) is a modification of 
$\ell_{\rm e}^{\rm OSF}$ multiplied by a factor  
but the valid region given by AA theory~\cite{ariel03b} 
is incompatible with that of OSF theory. 
OSF theory is held when the screening length $\kappa^{-1}$ is much 
smaller than the bare persistence length~\cite{everaers02}.
Therefore, AA theory may be true in the region $\kappa\gg 1/b$.  
As shown in the figure, $\ell_{\rm e}$ for vanishing $\sigma_{+4}$ is well 
described by the right branch of the curve of Eq.~(\ref{le_ariel}), 
which seems to support this conjecture. 
This topic deserves a further analysis in the future.

\subsection{Potential of mean force}
\label{Sec_pmf}
Experiments and simulations have shown that 
like-charged macroions can attract each other in a 
solution~\cite{gelbart00,angelini03,molnar05}. 
It is hence relevant to know under which conditions the like-charge attraction 
takes place between polyelectrolytes.
To answer this question, we investigate ``potential of mean force'' 
in this section. 
Potential of mean force describes the effective interaction between 
two particles (molecules) in a medium, which includes two contributions:
the direct interaction between two particles (molecules) and 
the indirect interaction via other particles (molecules) and the medium. 
This quantity can be calculated by simulations:
$W_{\alpha\beta}(r)=-k_{\rm B}T \ln g_{\alpha\beta}(r)$
where $g_{\alpha\beta}(r)$ is the radial distribution function between 
particles of species $\alpha$ and $\beta$. 
$g_{\alpha\beta}(r)$ is computed by 
\begin{equation}
g_{\alpha\beta}(r)=\frac{V \langle H_{\alpha\beta}(r)\rangle}
{N_{\alpha}N_{\beta} (4\pi r^2 \delta r)}
\end{equation}
where $N_{\alpha}$ and $N_{\beta}$ are the number of $\alpha$-particles 
and that of $\beta$-particles, respectively,  
$V$ is the volume of the simulation box, and  
$H_{\alpha\beta}(r)=\sum_{i=1}^{N_{\alpha}} h_{i\beta}(r)$ with
$h_{i\beta}(r)$ denoting the number of $\beta$-particles 
lying in a spherical shell, centered at the $\alpha$-particle $i$,
of radius $r$ and shell thickness $\delta r$.

\begin{figure}[htbp] 
\begin{center}
\includegraphics[angle=270,width=0.40\textwidth]{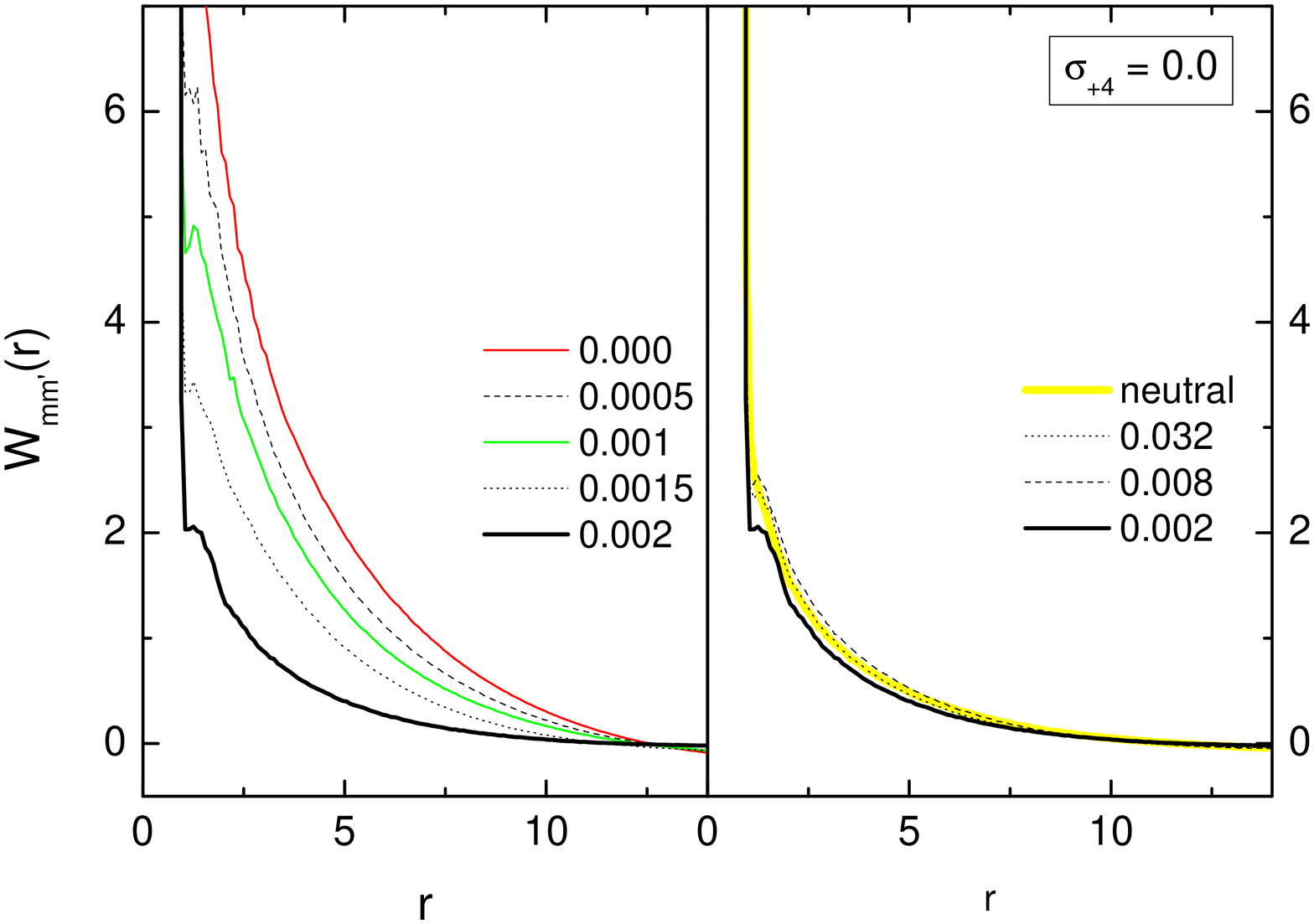}
\includegraphics[angle=270,width=0.40\textwidth]{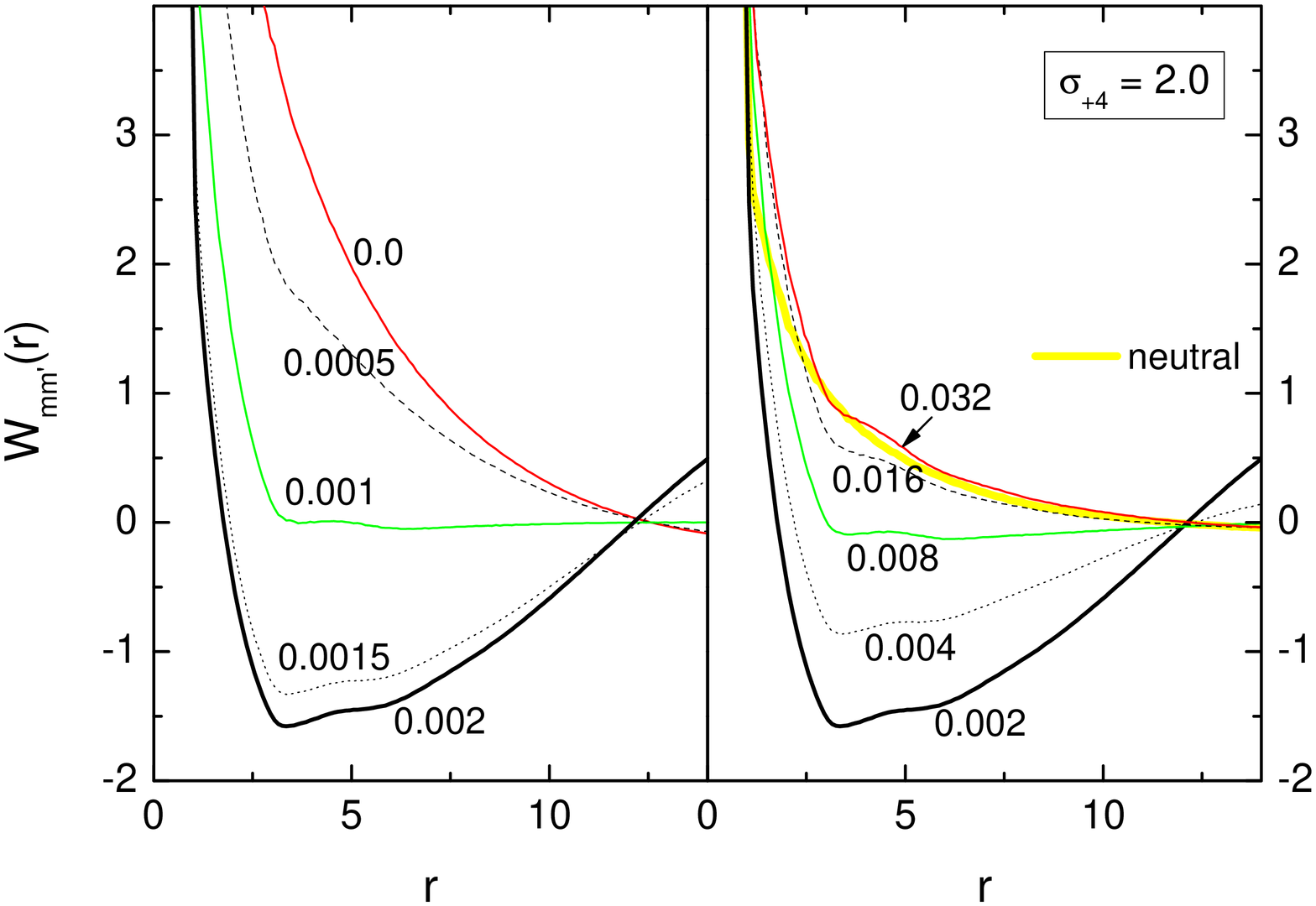}
\includegraphics[angle=270,width=0.40\textwidth]{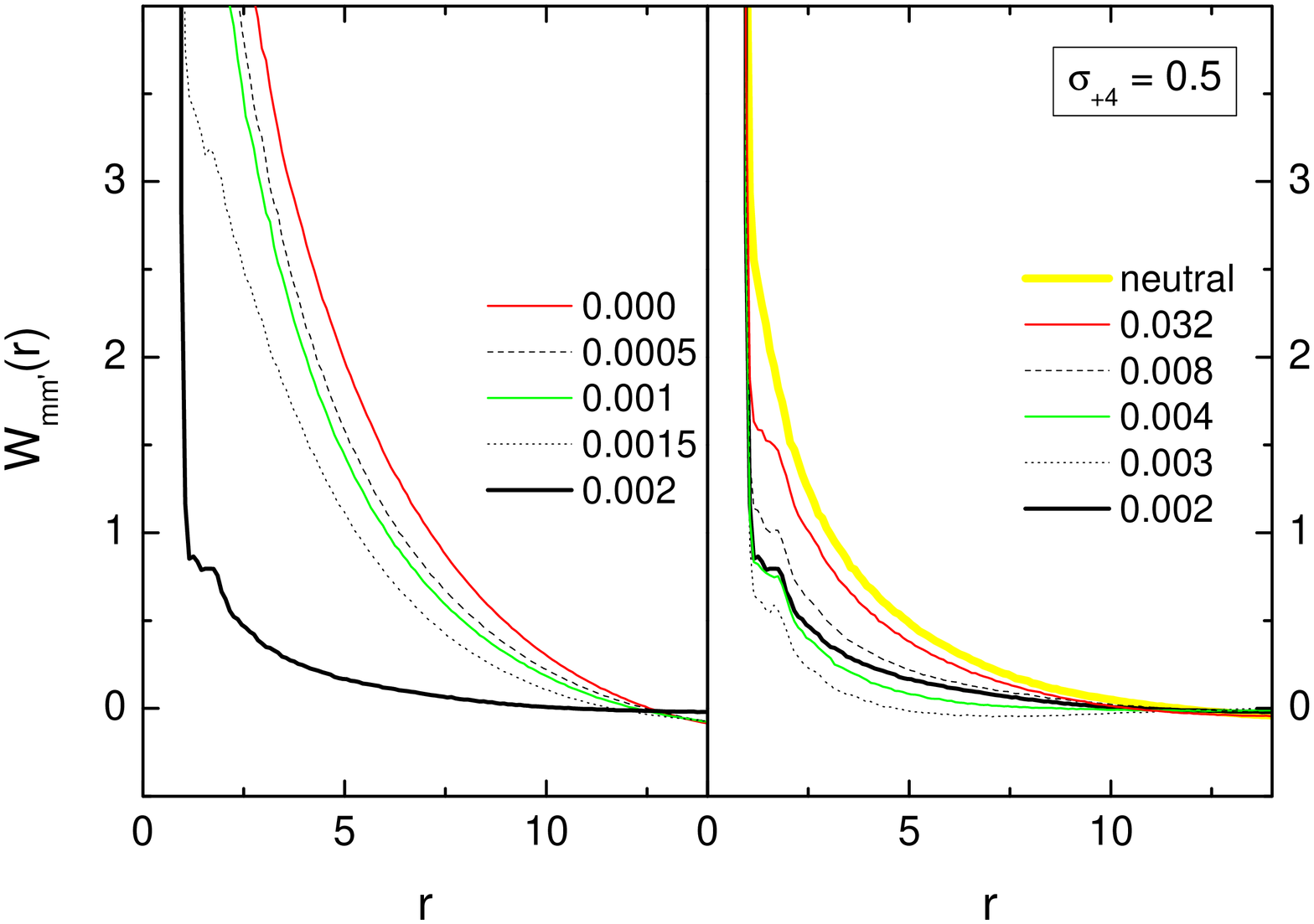}
\includegraphics[angle=270,width=0.40\textwidth]{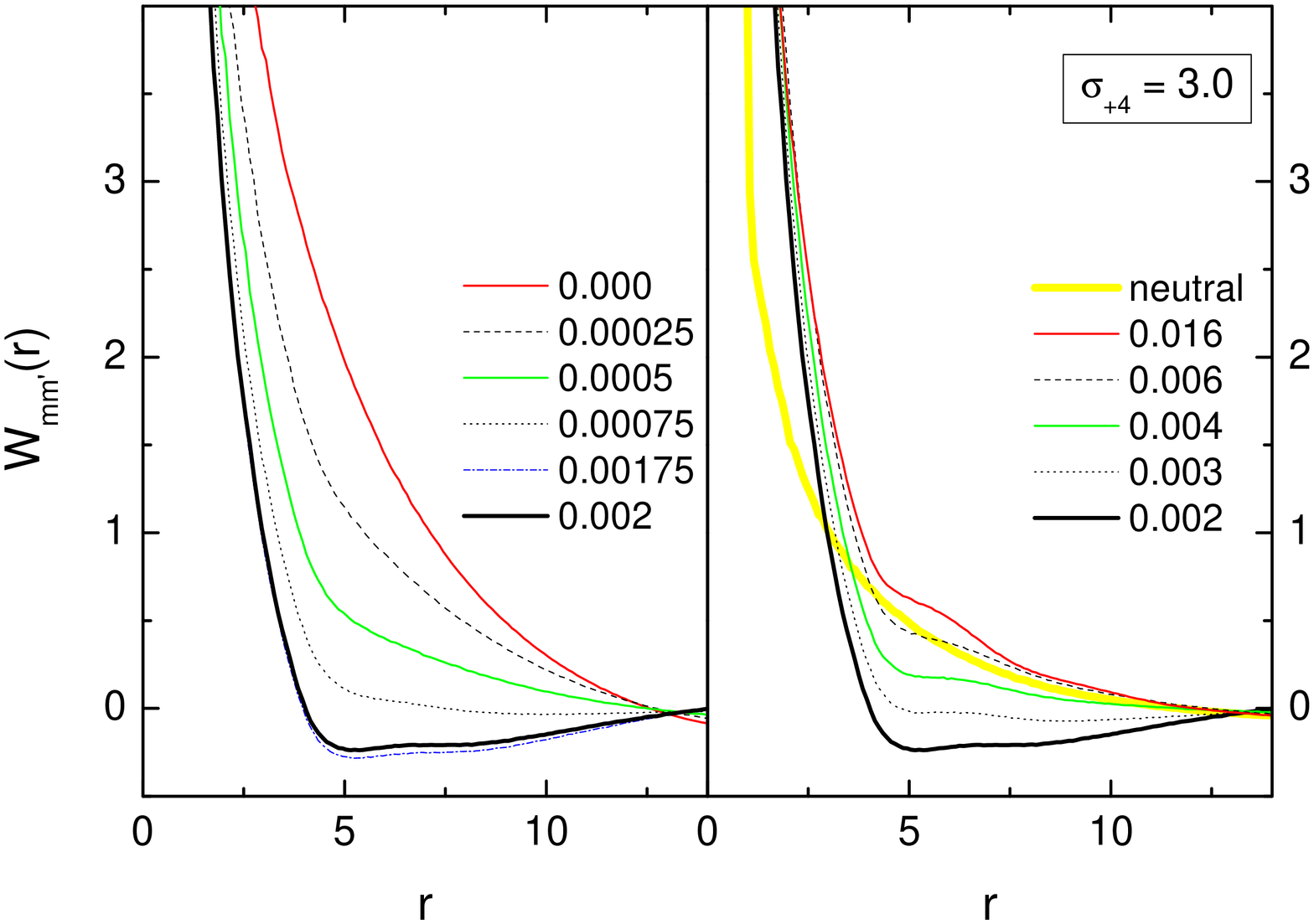}
\includegraphics[angle=270,width=0.40\textwidth]{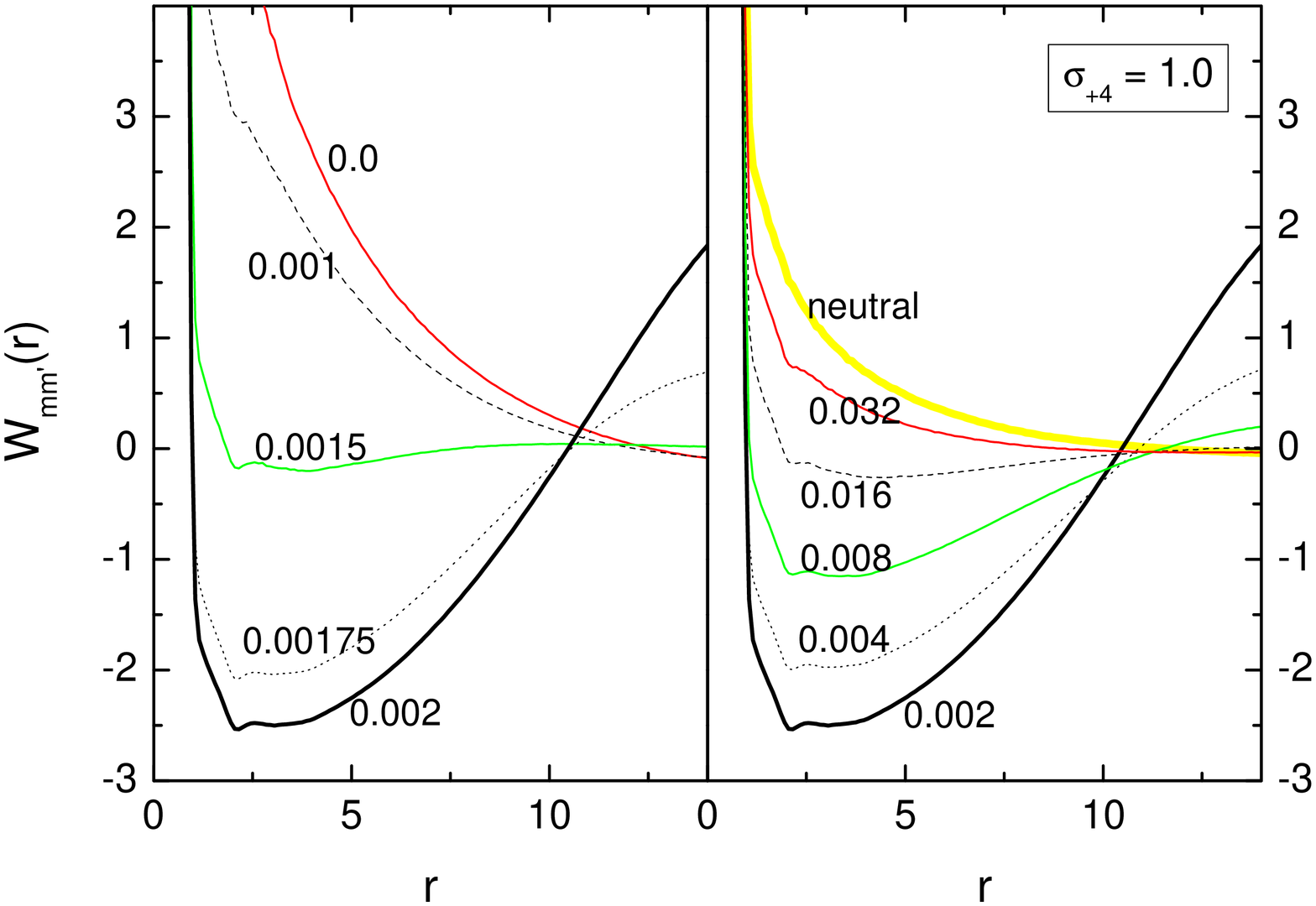}
\includegraphics[angle=270,width=0.40\textwidth]{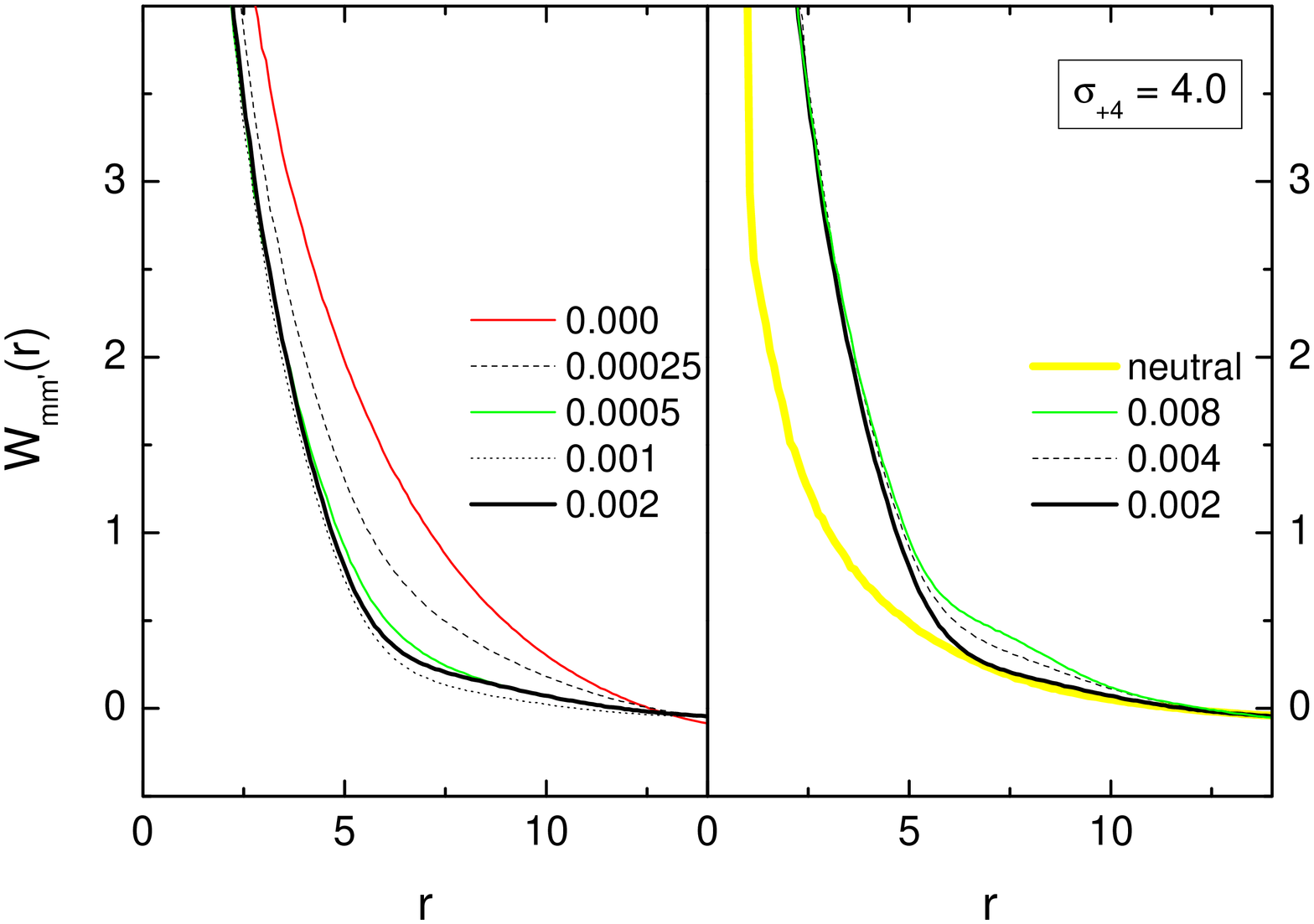}
\caption{$W_{\rm mm'}(r)$ for $\sigma_{+4}=0.0$, 0.5, 1.0, 2.0, 3.0 and 
4.0, shown, in orders, from top to bottom and from left to right, in six plots.
Each plot includes several $W_{\rm mm'}(r)$ at different $C_{\rm s}$.
The value of  $C_{\rm s}$ is indicated in the plot or near the corresponding curve.
$W_{\rm mm'}(r)$ for the neutral polymer system is depicted as reference and
marked by ``neutral''. The unit of ordinate axis is $k_{\rm B} T$.}
\label{PMF11P4N48txxx}
\end{center}
\end{figure}

Because our system is composed of only four chains, it is not easy to reduce 
statistical error if we directly calculate the potential of mean force 
between chains.
We, therefore, calculated the potential of mean force between monomers on different 
chains, which also provides the information about the chain interaction. 
We denote this quantity $W_{\rm mm'}(r)$ and the results are presented in Fig.~\ref{PMF11P4N48txxx} 
which contains six plots, from top to bottom and from left to right, corresponding, in orders, to 
the cases $\sigma_{+4}=0.0$, 0.5, 1.0, 2.0, 3.0 and 4.0.
$W_{\rm mm'}(r)$ for the referenced neutral polymers is also depicted in the figure.
In a salt-free solution, $W_{\rm mm'}(r)$ displays a repulsive interaction, and  
is stronger than that for the neutral polymers owing to the electrostatic repulsion 
between monomers.
This can be seen by comparing the range of repulsion estimated by equaling $W_{\rm mm'}(r)$ to 
the thermal energy $k_{\rm B} T$. 
The range of repulsion is at $r=7.1$ for the polyelectrolytes and at $r=3.0$ for the neutral polymers.
Upon addition of salt, $W_{\rm mm'}(r)$ responses in two distinct ways, which depends
on the ion size.
The first one is applied for small or large tetravalent counterions,
such as $\sigma_{+4}=0.0$, 0.5, or 4.0. 
In these cases, $W_{\rm mm'}(r)$ is purely repulsive in the course of addition of salt.
Hence, the monomers on different chains repel between themselves and the system exhibits no 
like-charge attraction.
The second way is happened for an intermediate size such as $\sigma_{+4}=1.0$, 2.0, or 3.0. 
At this moment, $W_{\rm mm'}(r)$ displays an attractive well 
in a mid-salt region.
Therefore, it appears like-charge attraction between chains.
In the first case, the range of repulsion of $W_{\rm mm'}(r)$ decreases to a minimum value 
nearly at $C_{\rm s}=0.002$.  Further increase of $C_{\rm s}$ either leaves $W_{\rm mm'}(r)$ 
unchanged (for vanishing $\sigma_{+4}$ and for $\sigma_{+4}=4.0$) 
or increases the range of repulsion of $W_{\rm mm'}(r)$ (for $\sigma_{+4}=0.5$). 
We will call $W_{\rm mm'}(r)$ at $C_{\rm s}=0.002$ ``the curve at equivalence'',    
which separates two behaviors of evolution of $W_{\rm mm'}(r)$ 
as $C_{\rm s}$ is increased or decreased from $C_{\rm s}^{*}$.
We noticed that $W_{\rm mm'}(r)$ tends toward the referenced curve
of the neutral polymers at high salt concentrations except for the deviation 
happened at small $r$ due to the excluded volume of large ions. 
For vanishing $\sigma_{+4}$, the curve at equivalence is very close to the reference curve 
and $W_{\rm mm'}(r)$ is roughly unchanged while $C_{\rm s}>0.002$.
For $\sigma_{+4}=0.5$, the curve at equivalence  
lies below the referenced curve; $W_{\rm mm'}(r)$ is bounded between 
the two curves and increases with $C_{\rm s}$.
And in the second case, $W_{\rm mm'}(r)$ displays a potential well 
located near $r=\sigma_{\rm m}+\sigma_{+4}$. 
It suggests an attraction mediated by a tetravalent counterion. 
The depth of the well can be more profound than $k_{\rm B} T$.
Therefore, a stable bound state can be formed between monomers on different chains and
it is a prerequisite to occur salt-induced phase separation.
The maximum depth of the well is happened while $C_{\rm s}$ is $C_{\rm s}^{*}$. 
In general, we found that $W_{\rm mm'}(r)$ is bounded between the salt-free curve and
the curve at equivalence while $C_{\rm s}<0.002$, and  
between the latter curve and the referenced neutral curve while $C_{\rm s}>0.002$.

\begin{figure}[htbp] 
\begin{center}
\includegraphics[angle=270,width=0.40\textwidth]{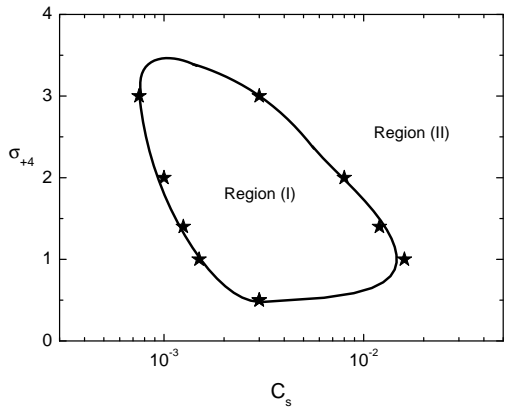}
\caption{Two distinct regions in the ($C_{\rm s}$, $\sigma_{+4}$)-variable space. 
In Region (I), $W_{\rm mm'}(r)$ shows an attractive well and in Region (II), it is purely repulsive.}
\label{PMF11P4N48_boundary.eps}

\end{center}
\end{figure}

We have shown that the occurrence of the like-charge attraction 
depends on both salt concentration and ion size. 
In Fig.~\ref{PMF11P4N48_boundary.eps} we present, in the variable space,
the region where $W_{\rm mm'}(r)$ shows an attractive well. 
In the figure, the data points (symbolized by star) denote the 
salt concentrations for a given $\sigma_{+4}$, at which 
a purely repulsive $W_{\rm mm'}(r)$ turns to show an attractive well 
and \textit{vice versa}, and the closed solid curve is sketched to help readers 
to separate the two regions. 
It clearly shows that an effective attraction appears only when 
$\sigma_{+4}$ is neither too small nor too big and salt concentration is intermediate
around the equivalence concentration.  
We mention that the occurrence of an attractive region in the potential of mean force
is not sufficient to have a phase separation of the system.
Phase transitions in charged colloid systems have been largely discussed in literatures 
but there have been relatively few studies of phase transitions in charged chain systems 
(see Refs.~\cite{taboada-serrano05} and \cite{panagiotopoulos05} and references therein). 

\begin{figure}[htbp] 
\begin{center}
\includegraphics[angle=0,width=0.40\textwidth]{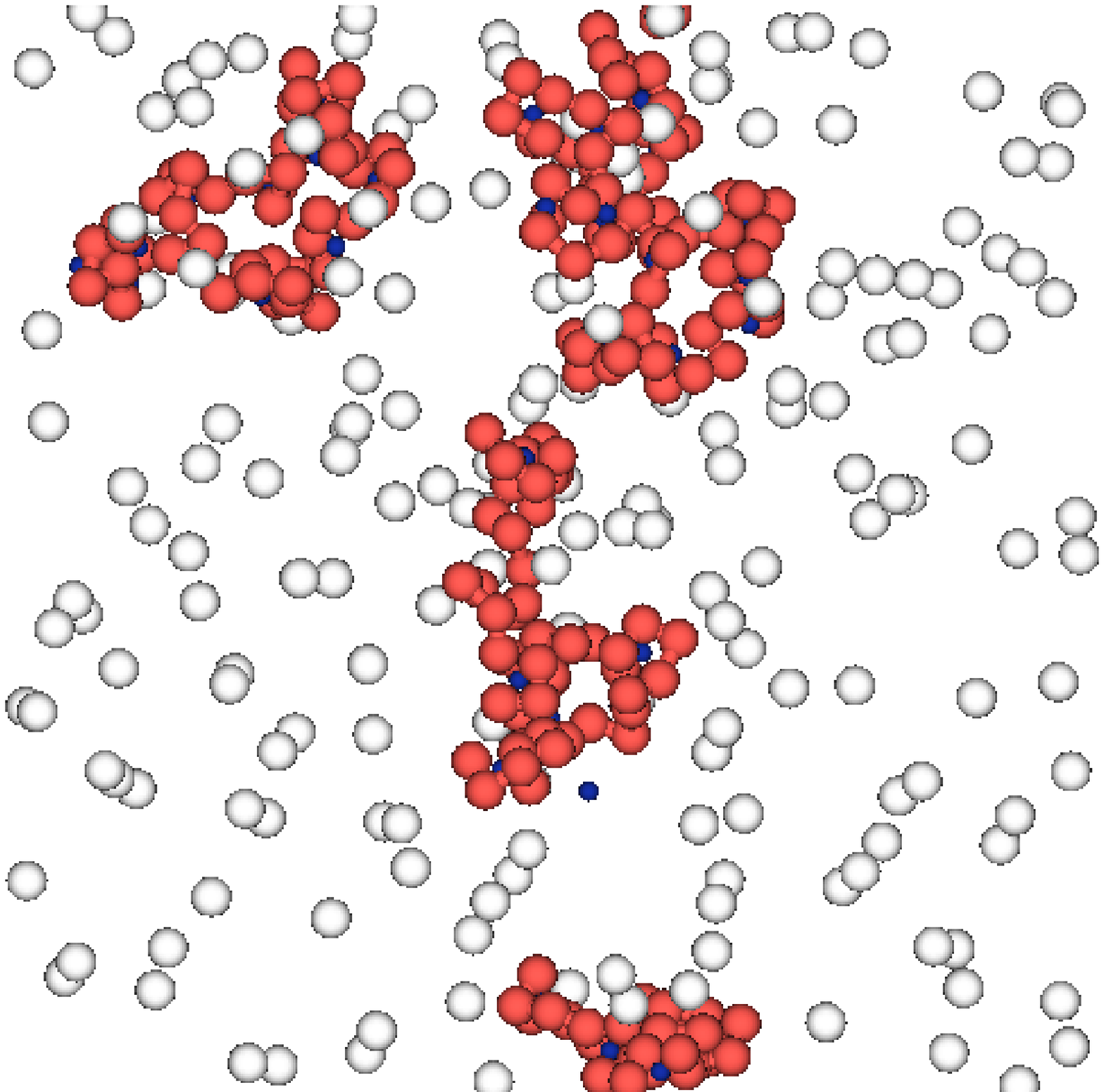}
\includegraphics[angle=0,width=0.40\textwidth]{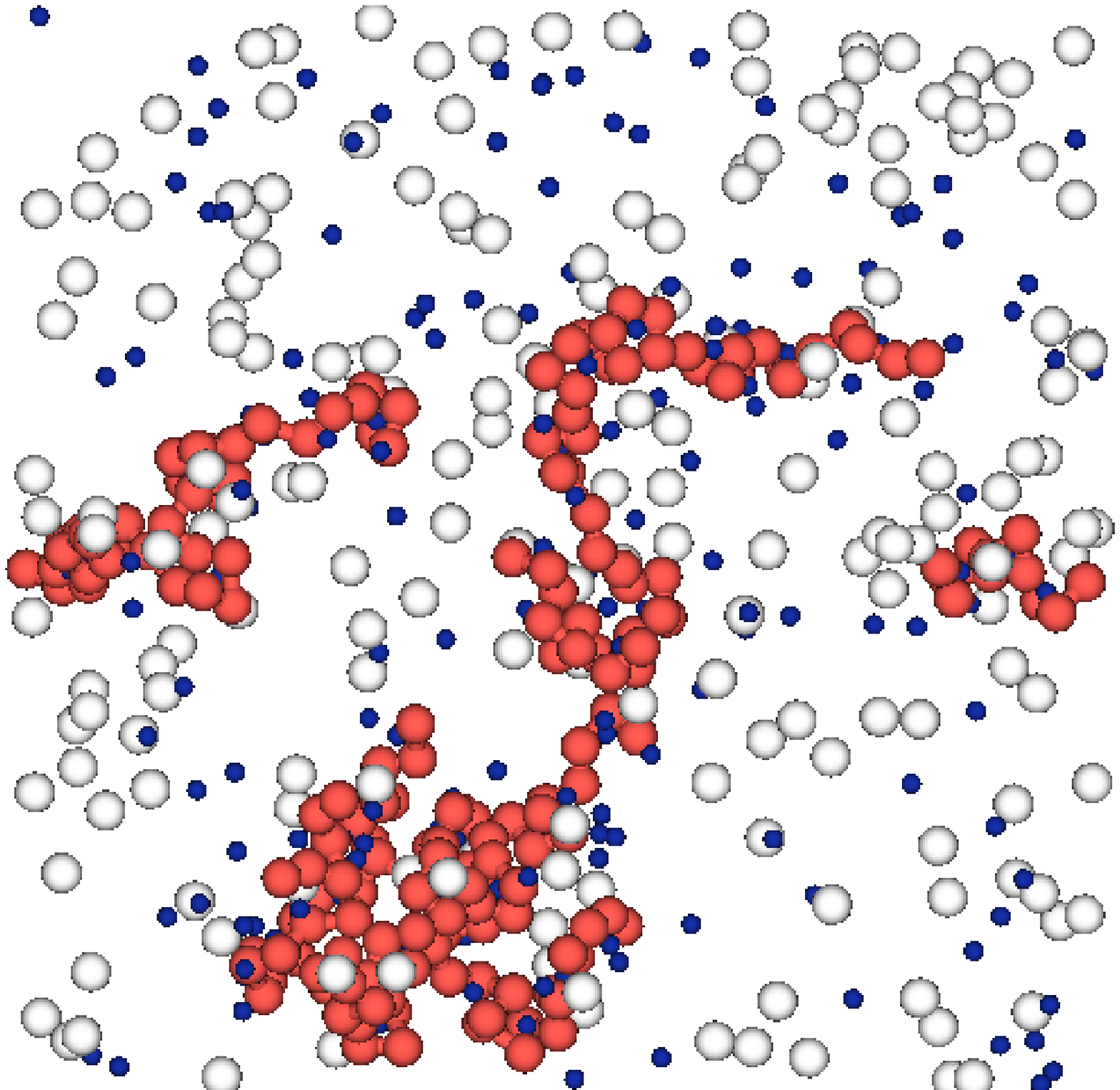}
\end{center}
\center{(a)}
\begin{center}
\includegraphics[angle=0,width=0.40\textwidth]{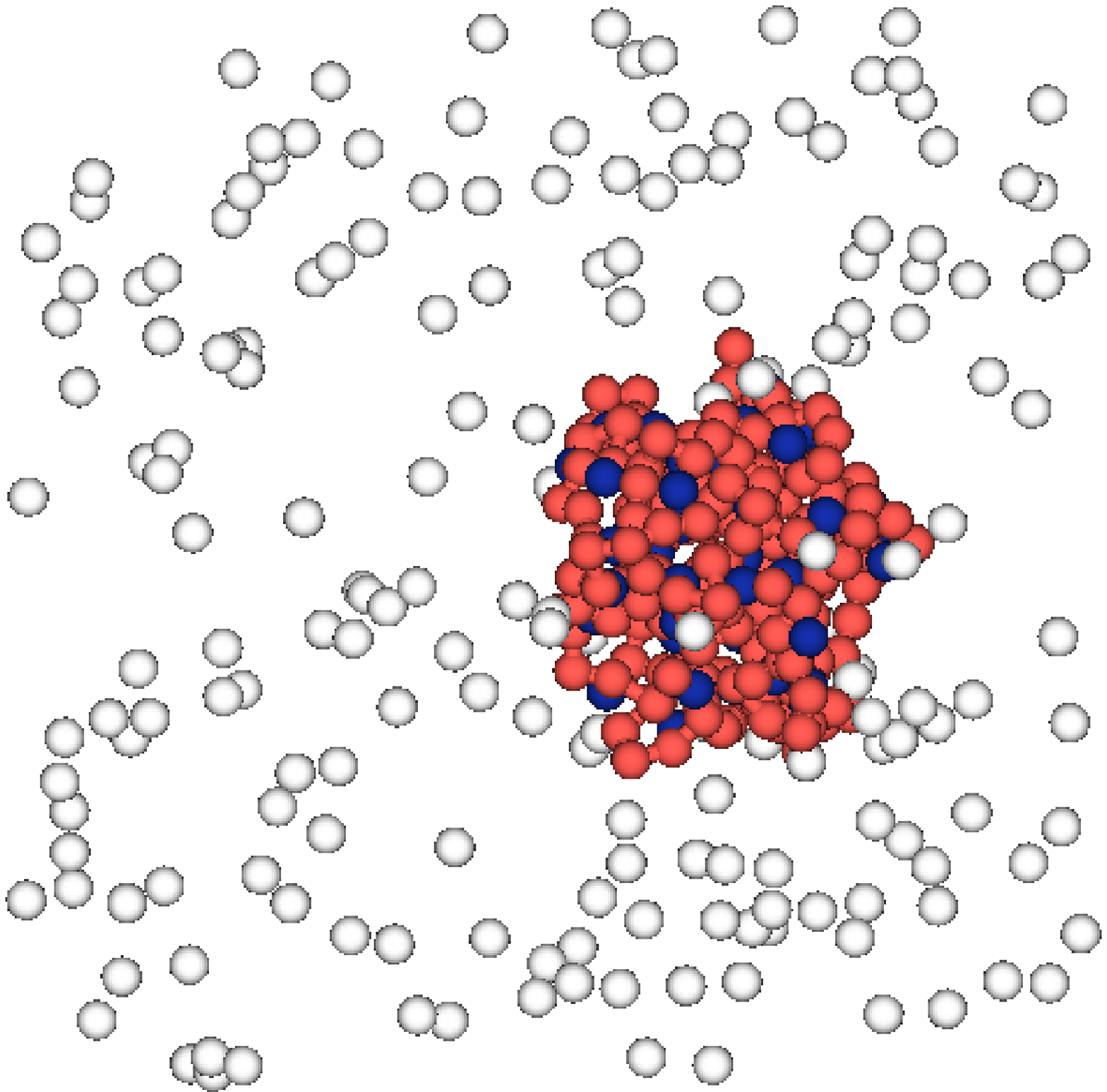}
\includegraphics[angle=0,width=0.40\textwidth]{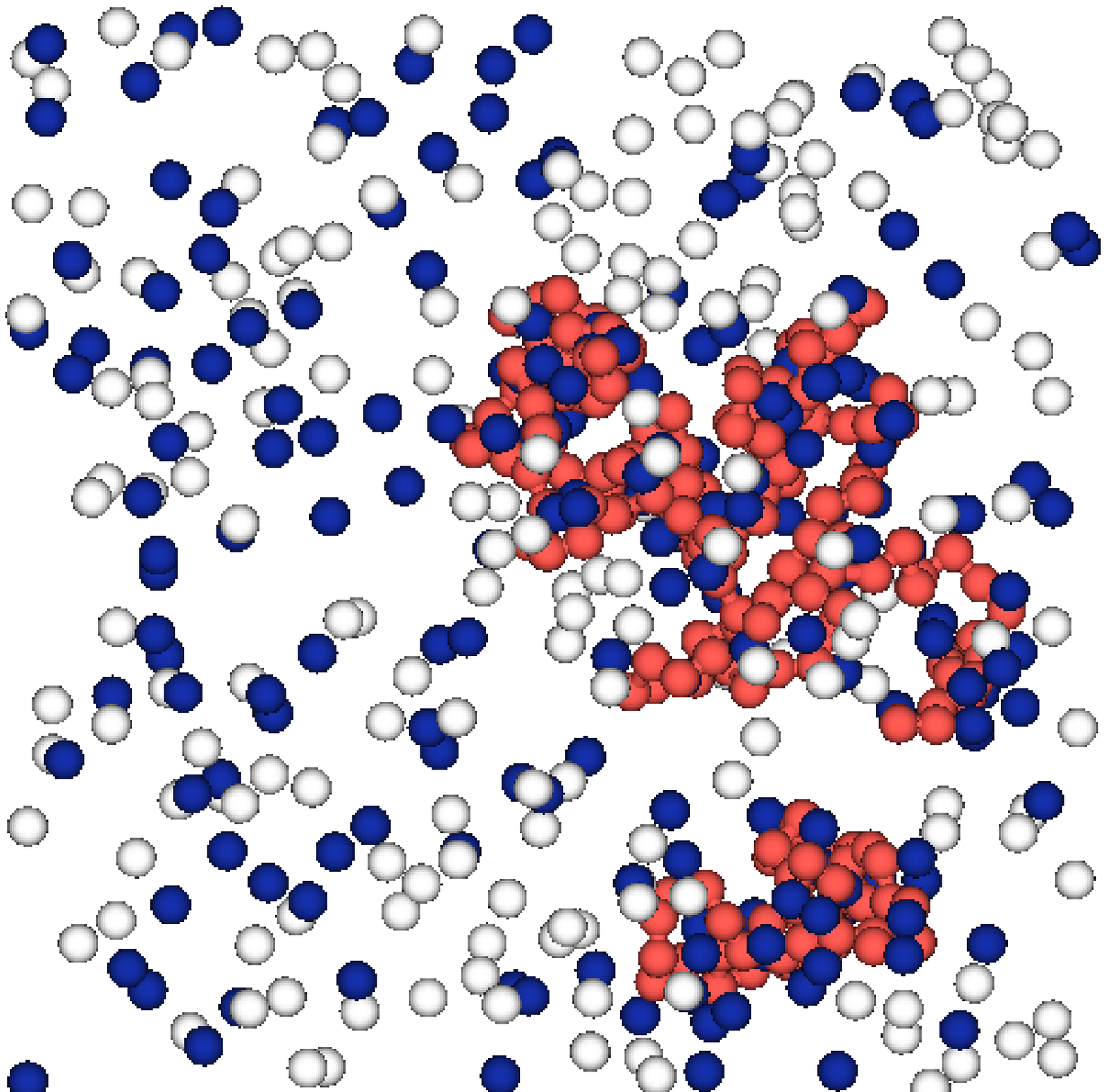}
\end{center}
\center{(b)}
\begin{center}
\includegraphics[angle=0,width=0.40\textwidth]{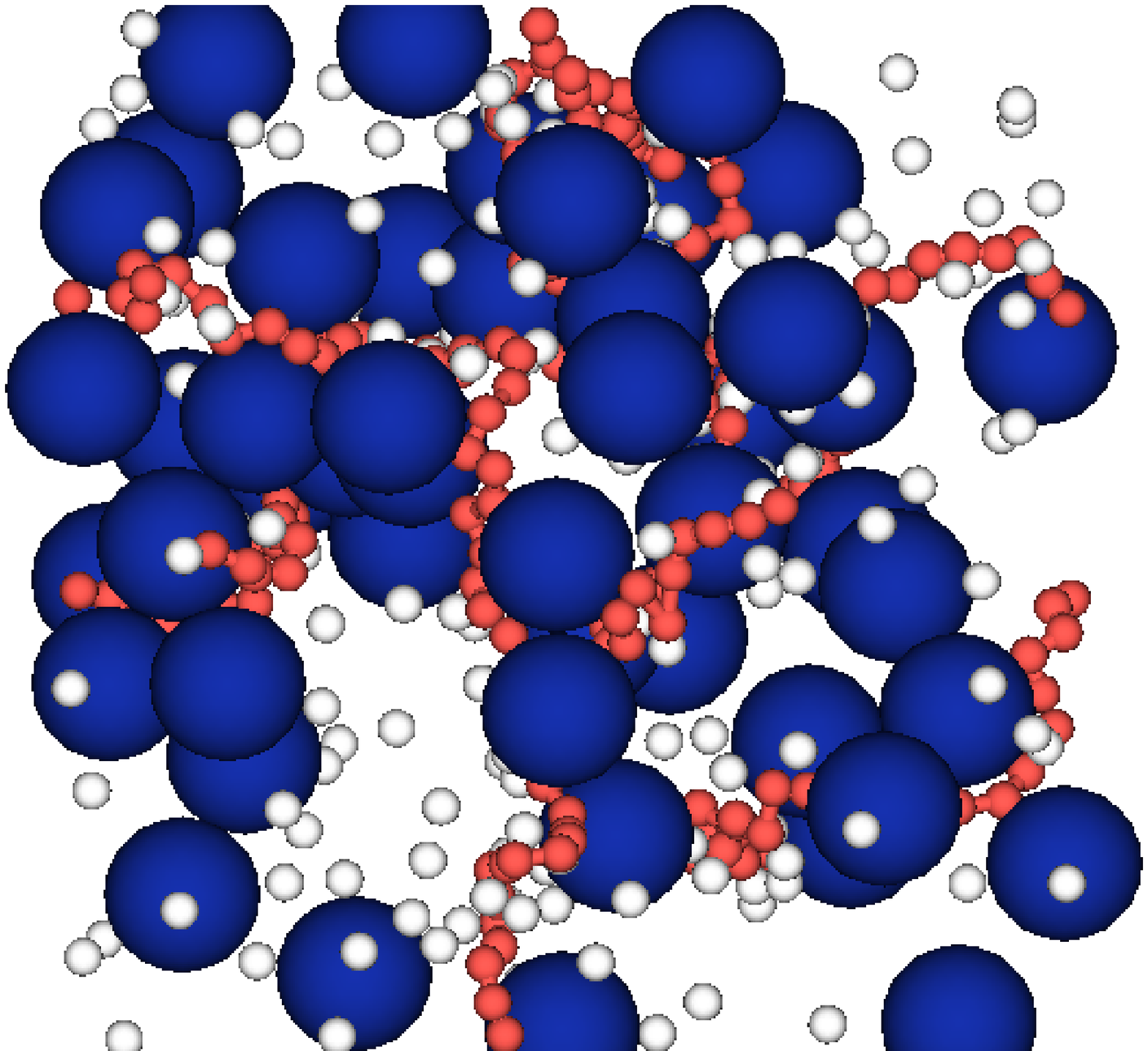}
\includegraphics[angle=0,width=0.40\textwidth]{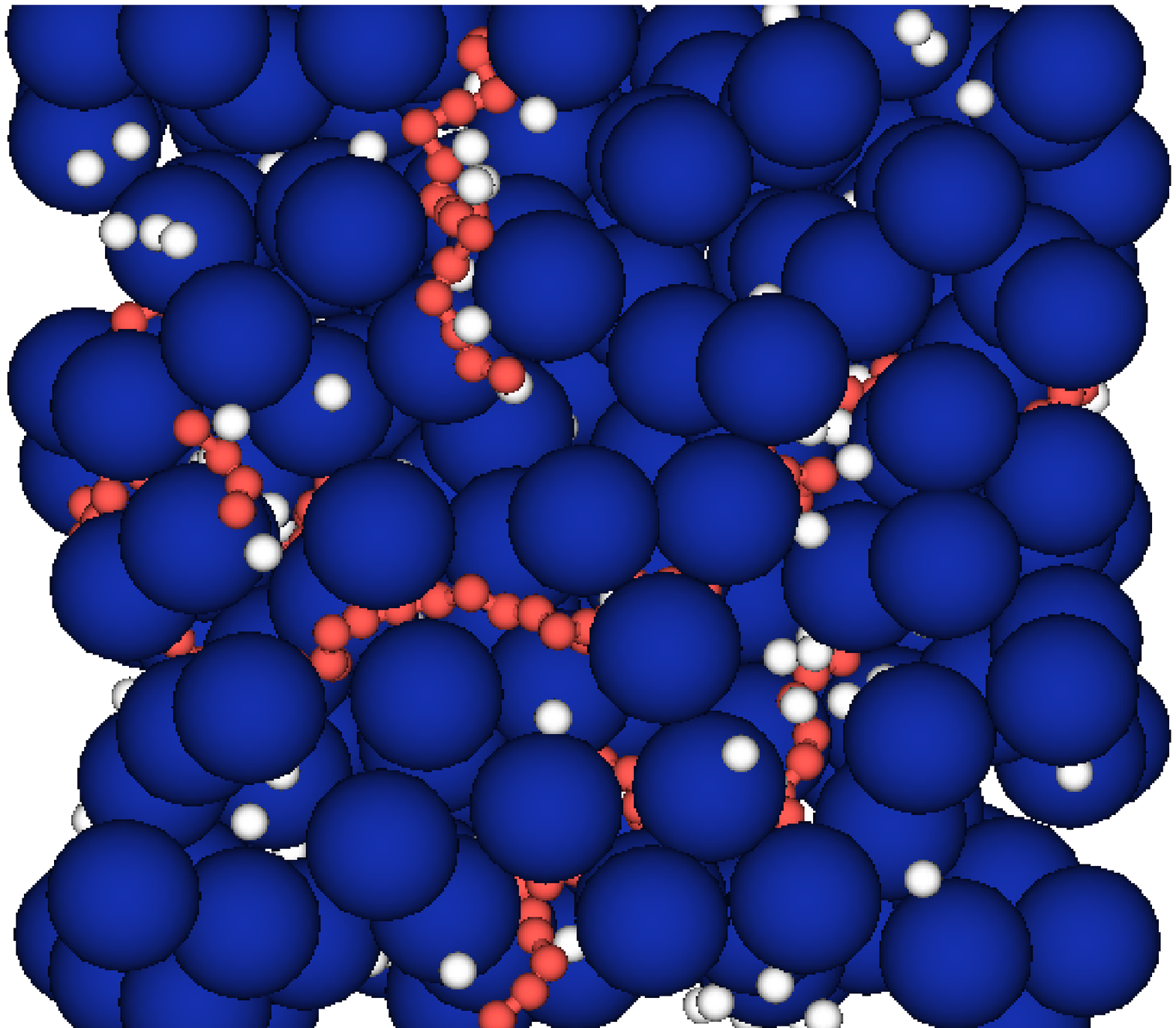}
\center{(c)}
\caption{(Color online) Snapshots of simulations for
(a) $\sigma_{+4}=0.5$, (b)  $\sigma_{+4}=1.0$, and (c) 
 $\sigma_{+4}=4.0$, at $C_{\rm s}=0.002$ (left) and 
$C_{\rm s}=0.008$ (right).
In the pictures, the bead-spring chains are represented
in gray color (red color online) and the monovalent counterions 
in light one (white color online).
The tetravalent counterions are represented
in dark-colored (blue-colored online) particles whereas the  
coions are not plotted for the clarity of the pictures.
In the snapshots, a chain may appear more than one fragment
because of periodic boundary condition.}
\label{snapshots}
\end{center}
\end{figure}

Fig.~\ref{snapshots} shows snapshots of the simulations for
$\sigma_{+4}=0.5$, 1.0 and 4.0 at two salt concentrations 
$C_{\rm s}=0.002$ and $C_{\rm s}=0.008$.
At $C_{\rm s}=0.002$, we saw that the polyelectrolytes favor to be apart 
from each other for both of the cases $\sigma_{+4}=0.5$ and $\sigma_{+4}=4.0$. 
However, delicate difference was witnessed that the previous case displays 
a single-chain collapse but the latter one does not.
For $\sigma_{+4}=1.0$, the chains favor to aggregate together,
in accordance with the results of Fig.~\ref{PMF11P4N48txxx} which 
show the occurrence of an attraction between the negatively-charged polymers.
This attraction is mediated by counterions, mainly the tetravalent counterions,
since almost all the tetravalent counterions are condensed on the polyelectrolytes. 
We point out that the condensation of tetravalent counterion
is not a sufficient condition to occur chain aggregation. 
It can be seen from the case $\sigma_{+4}=0.5$ in which  
most of the tetravalent counterions are condensed
but the chains does not aggregate. 
At the high salt concentration $C_{\rm s}=0.008$, 
the attraction between chains is reduced, in comparison with the cases at $C_{\rm s}=0.002$. 
The chains segregate or show tendency of segregation from each other. 
In the simulations, we observed that while $C_{\rm s}>0.002$, 
only part of the tetravalent counterions can condense onto the chains.

\subsection{Integrated charge distribution around a chain}
\label{Sec_IonDist}
In this section, we investigate the charge distribution around a
polyelectrolyte. It enables  to understand how ions distribute 
around a chain and provides useful information for theorists
to develop models.   
We define a worm-like tube of radius $r$ around a chain
to be a union of $N$ spheres of radius $r$ centered at each monomer 
on the chain.
We calculated the averaged total charge inside a worm-like tube,
including the charge of the ions and the one of the monomers. 
This quantity is called ``the integrated charge distribution'' and
is denoted by $Q(r)$.

\begin{figure}[htbp] 
\begin{center}
\includegraphics[angle=270,width=0.40\textwidth]{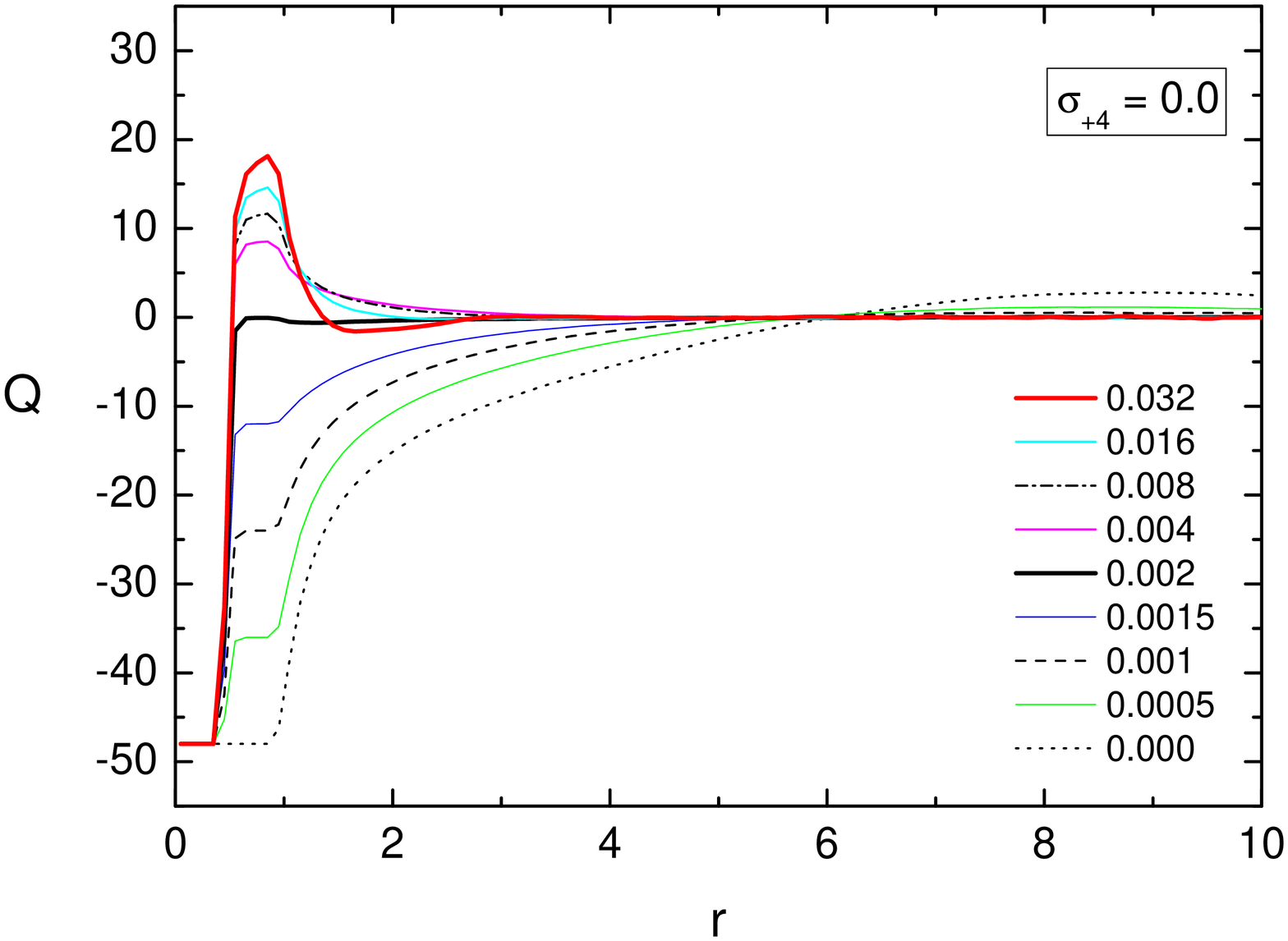}
\includegraphics[angle=270,width=0.40\textwidth]{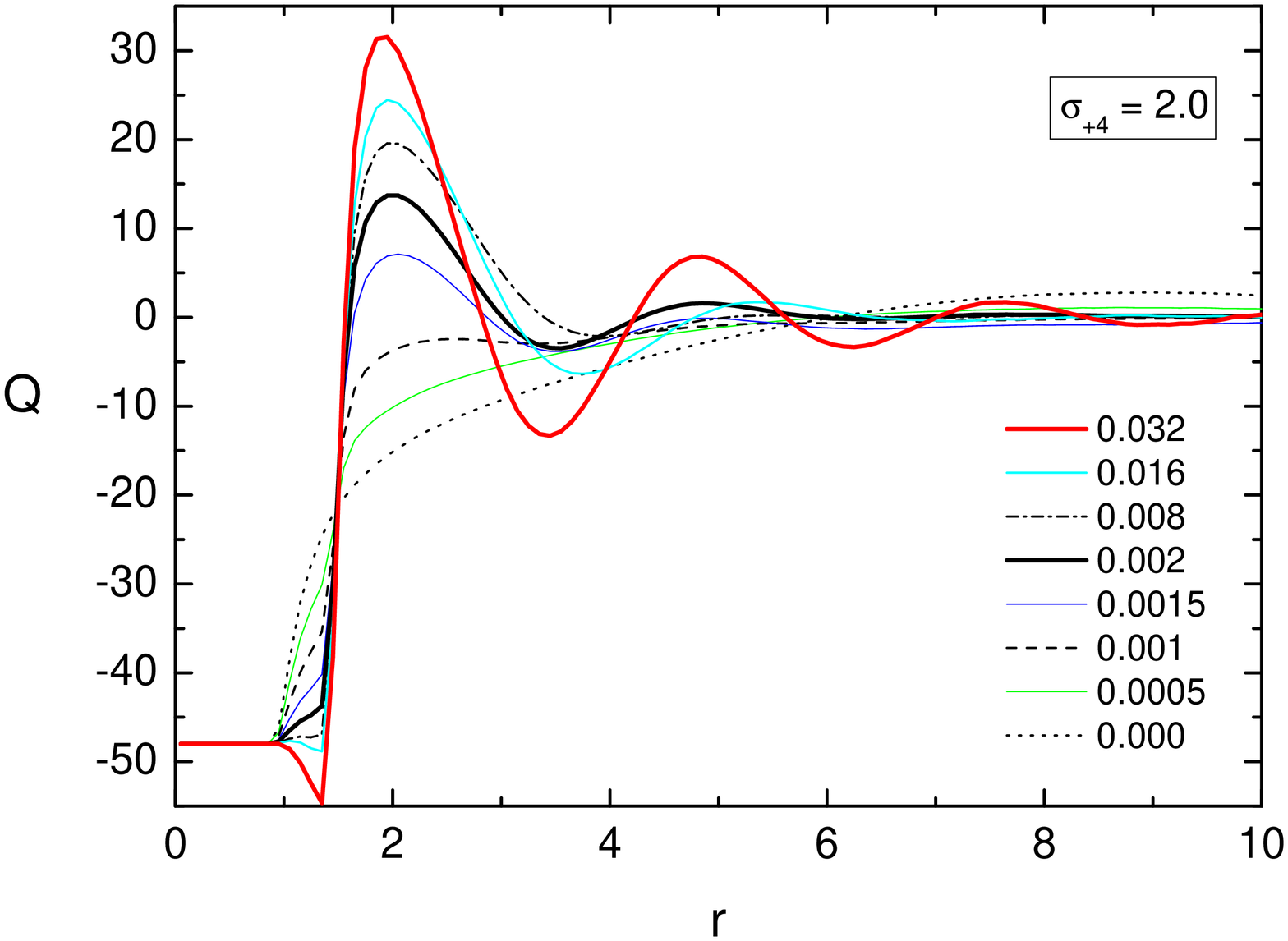}
\includegraphics[angle=270,width=0.40\textwidth]{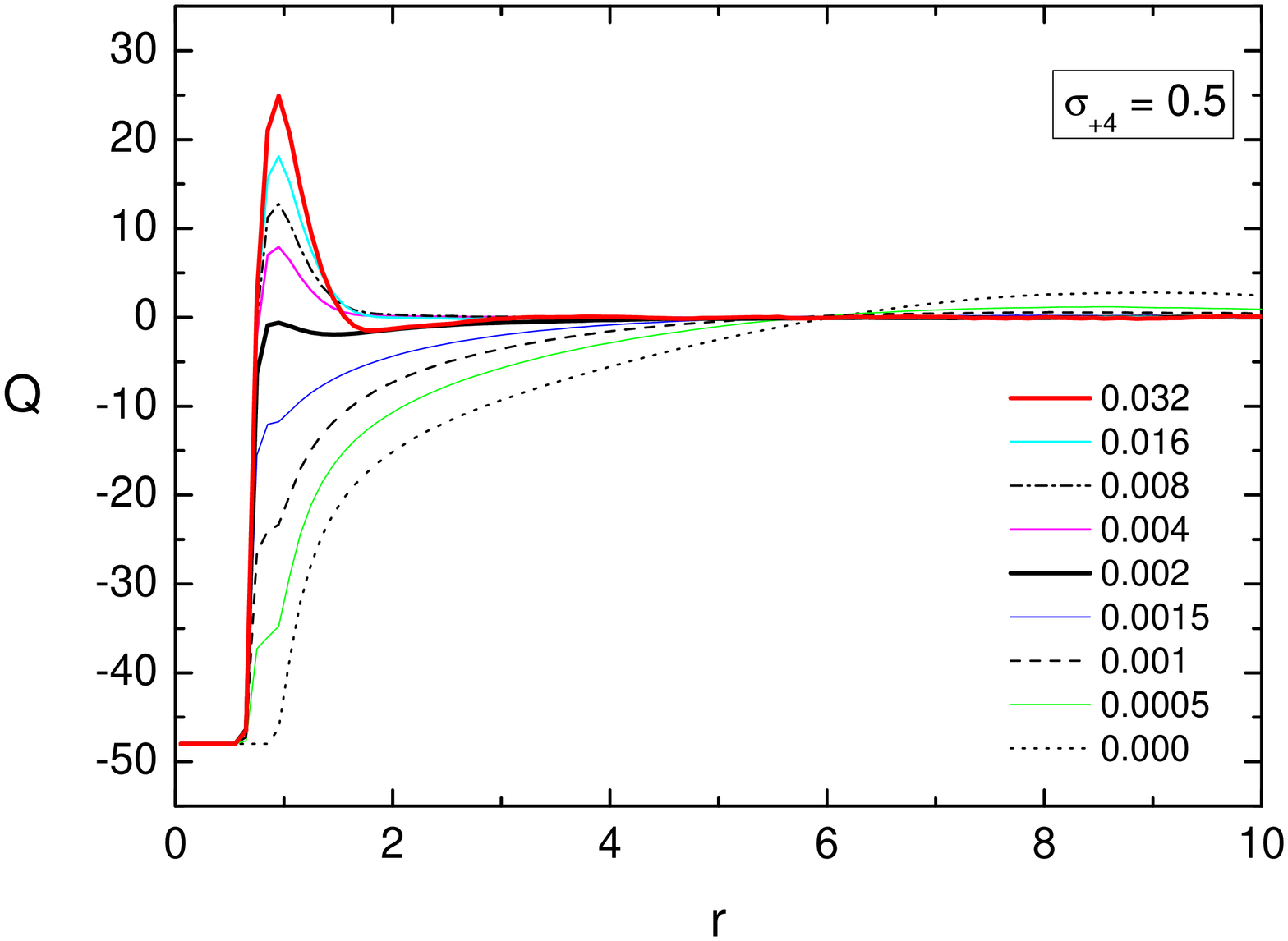}
\includegraphics[angle=270,width=0.40\textwidth]{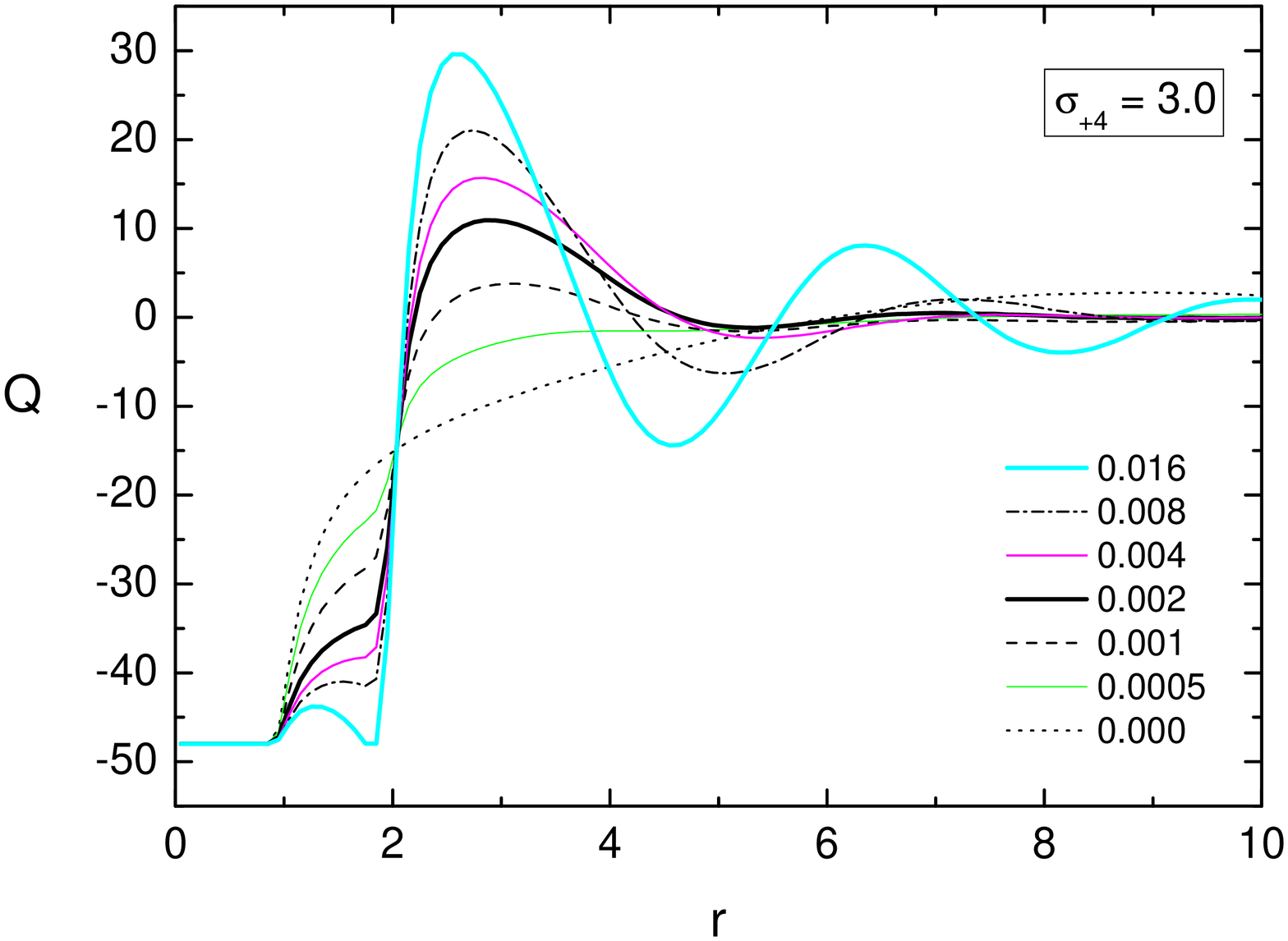}
\includegraphics[angle=270,width=0.40\textwidth]{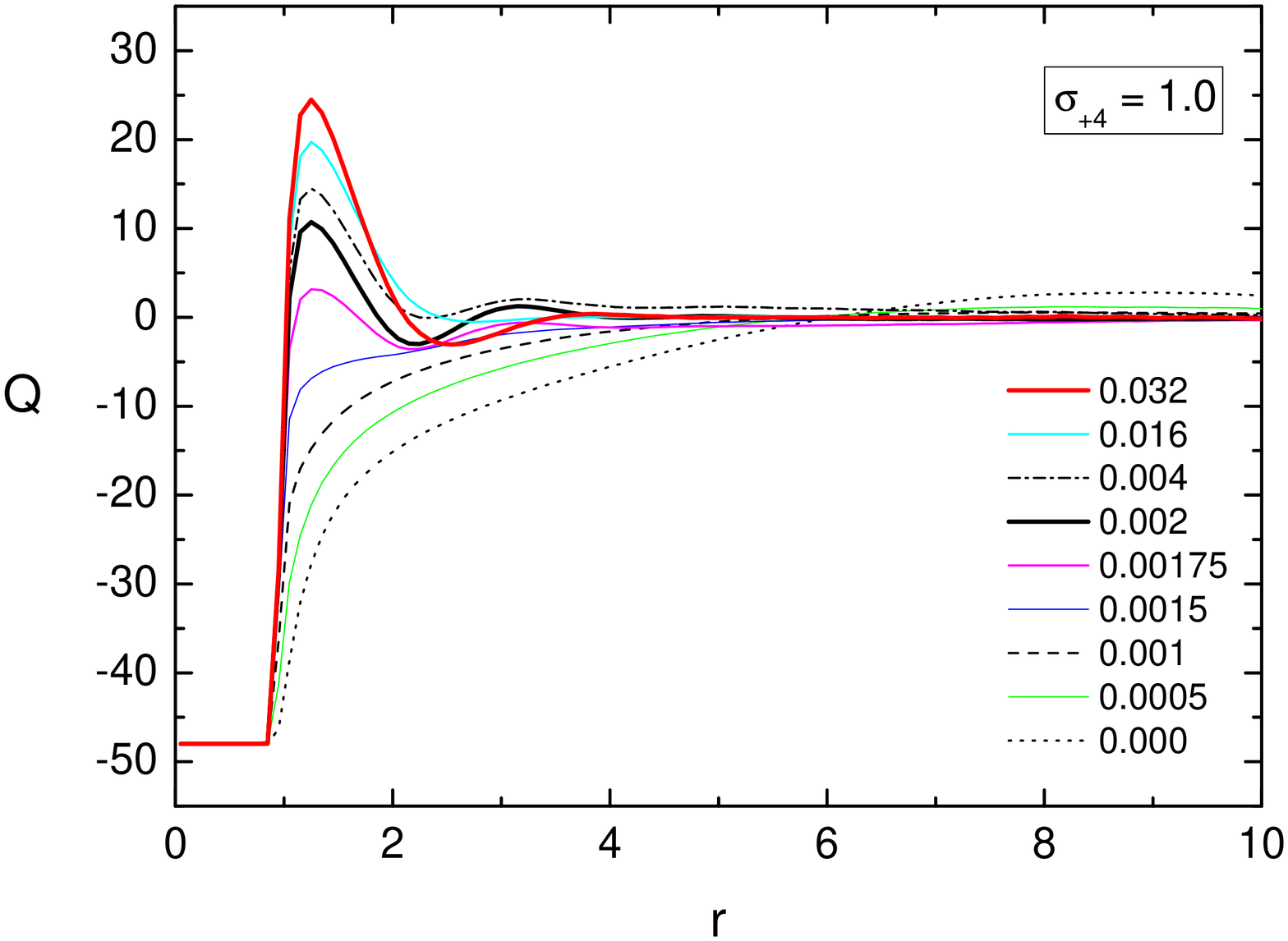}
\includegraphics[angle=270,width=0.40\textwidth]{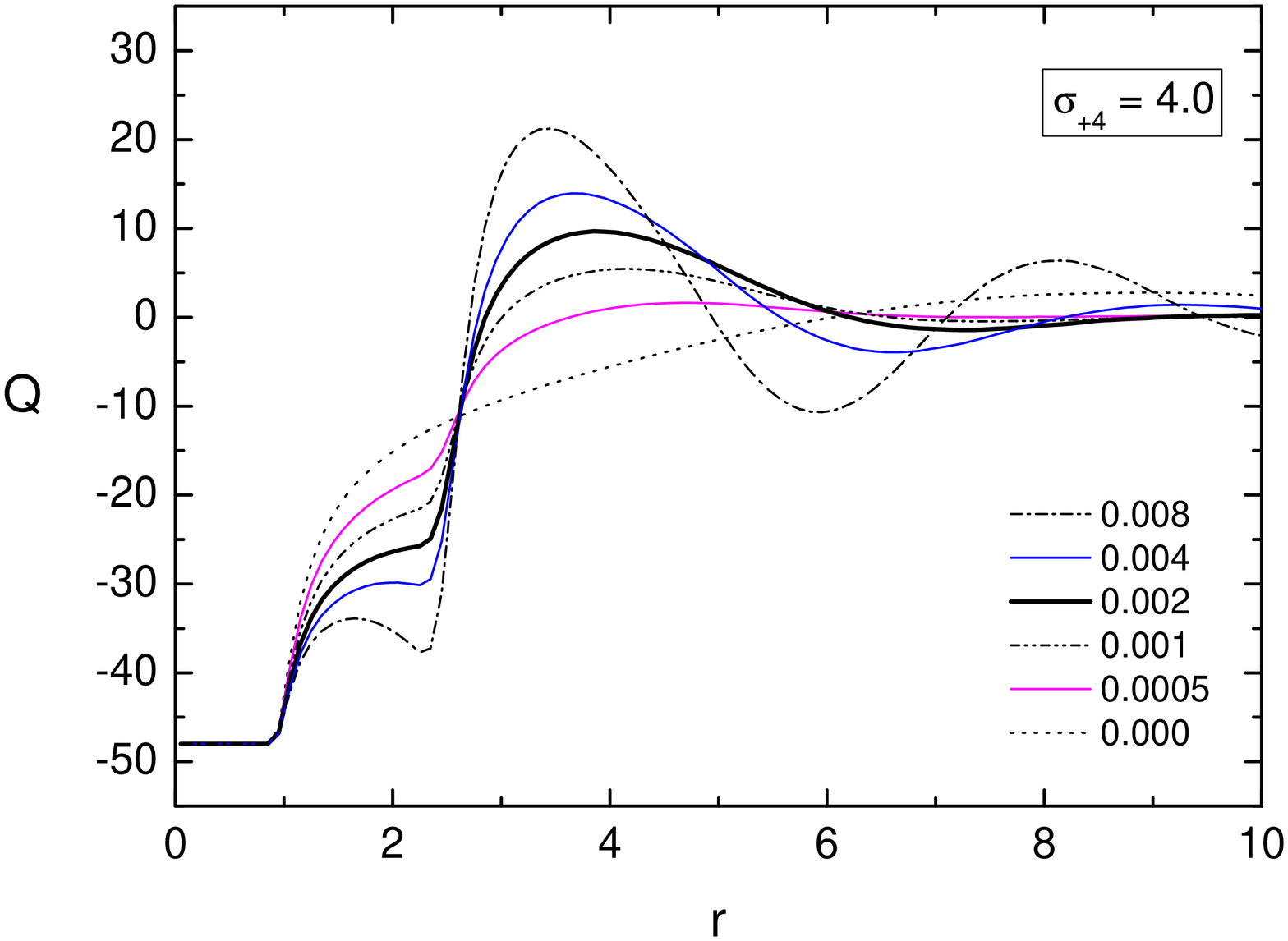}
\caption{$Q(r)$ at different  $C_{\rm s}$ for
$\sigma_{+4}$=0.0, 0.5, 1.0, 2.0, 3.0 and 4.0, shown, in orders, from top to bottom 
and from left to right, in six figures. The line arts used for different $C_{\rm s}$
are indicated at the right-bottom side of each figure.  }
\label{IonDist}
\end{center}
\end{figure}

We show in Fig.~\ref{IonDist} $Q(r)$ at various $C_{\rm s}$ 
for $\sigma_{+4}$ ranged between 0.0 and 4.0.
Focus firstly on the case with vanishing $\sigma_{+4}$.
In the region $r<0.5$, $Q(r)$ attains a value equal to the bare charge of a chain, $-48$.
A stair-like behavior then appears for $Q(r)$ in the region $0.5 <r< 1.0$ 
while $0.0< C_{\rm s}<0.002$.
Outside the region, $Q(r)$ monotonically increases and 
tends toward zero at large $r$ due to electroneutrality. 
Since $r<1.0$ is the depletion zone for the particles other 
than tetravalent counterions,  
the behavior of $Q(r)$ in this region is completely determined by the condensation of 
tetravalent counterions.
The stair-like behavior indicates a \textit{perfect} condensation
on the surface of the chain, owing to  the vanishing size of the tetravalent counterions.
For instance, at $C_{\rm s}=0.001$, we have 24 tetravalent counterions 
in the simulation box and, therefore, 6 condensed tetravalent counterions on a chain, 
in average.  A perfect condensation will lead to a value $-24$ for $Q(r)$ and 
this is exactly what we have observed in the figure. 
At the equivalence concentration, $Q(r)$ is zero while $r>0.5$.
If $C_{\rm s}$ is further increased ($C_{\rm s}>0.002$),
$Q(r)$ reveals a positive peak in the region  $0.5 < r < 1.0$, 
which indicates an overcharging, and, then, rapidly decays to zero as $r>1.0$.  
The hight of the peak suggests that only a portion of the tetravalent counterions 
are condensed on the chain.
For the case $\sigma_{+4}=0.5$, the region in which only the tetravalent counterions 
can intervene is narrowed to $0.75<r< 1.0$.
In the region,  the step of the stair-like $Q(r)$ becomes tilted while $C_{\rm s}<0.002$. 
If $C_{\rm s}>0.002$, $Q(r)$ displays a positive peak, which is more pronounced than that 
for vanishing $\sigma_{+4}$. 
For the case $\sigma_{+4}=1.0$,
an oscillatory behavior occurs at high $C_{\rm s}$.
Similar phenomena have been observed in other systems such as 
charged plate~\cite{greberg98}, colloid system~\cite{messina02}, 
and rigid polyelectrolyte~\cite{deserno02}.
This oscillatory behavior indicates a multi-layer organization of ions: 
in the close neighborhood of a polyelectrolyte, tetravalent counterions 
condense and form a positively-charged layer which overcharges the chain;
exterior to this layer, anionic particles intervene and
form a negatively-charged layer which overcompensates 
the charge inside and, hence, the integrated charge becomes negative, 
and so forth.  
Notice that the depletion zone around a chain for tetravalent counterions is $r<(1+\sigma_{+4})/2$.
It becomes larger than that for the other ions while $\sigma_{+4}>1.0$. 
In these cases, $Q(r)$ decreases with increasing $C_{\rm s}$ in the depletion zone,
but shows completely opposed behavior just outside the zone
apparently due to the condensation of tetravalent counterions. 
At very high $C_{\rm s}$ (for instance, at $C_{\rm s}=0.032$ 
for $\sigma_{+4}=2.0$), $Q(r)$ in the zone can even attain a value
smaller than the chain bare charge.
In this case, negative particles are the major composition of the first ion layer 
around a chain, which is counterintuitive.
The following three observations are worth to be noticed. 
(1) The salt concentration at which $Q(r)$ becomes (roughly) zero 
    in the region $r>(1+\sigma_{+4})/2$ decreases with increasing  $\sigma_{+4}$.
(2) The oscillatory behavior appears more markedly 
    for large $\sigma_{+4}$ than for small one. 
(3) At a given $C_{\rm s}$,  the larger the $\sigma_{+4}$,
    the higher the maximum peak of $Q(r)$ will be.  

\begin{figure}[htbp] 
\begin{center}
\includegraphics[angle=270,width=0.40\textwidth]{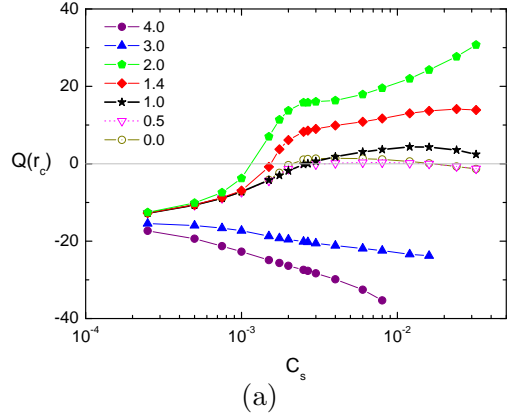}
\\
(a)
\\
\includegraphics[angle=270,width=0.40\textwidth]{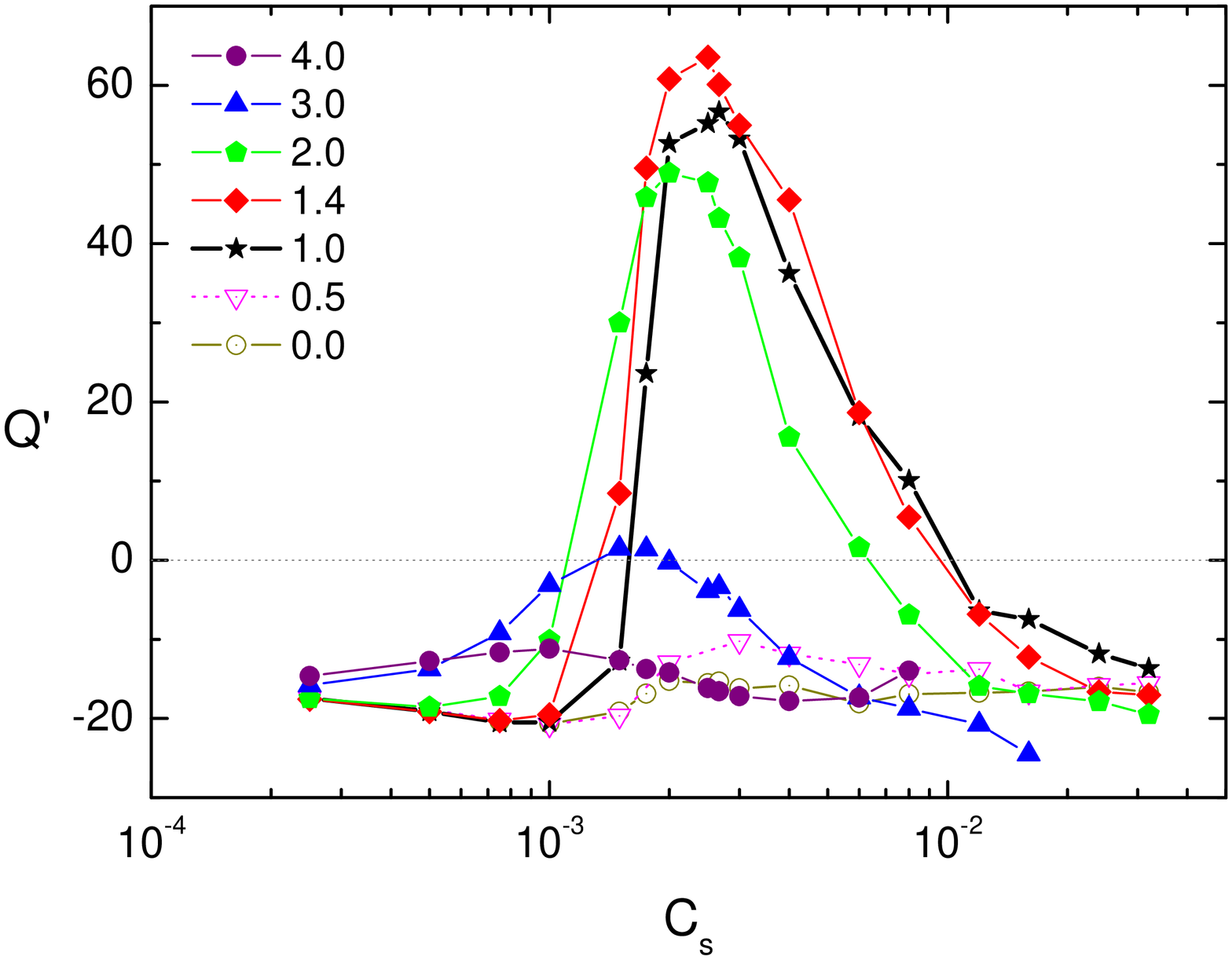}
\\
(b)
\caption{(a) $Q(r_c)$ as a function of $C_{\rm s}$ with $r_c=2.0$,
and (b) $Q'$ as a function of $C_{\rm s}$, 
for different $\sigma_{+4}$. 
The symbol used to denote each $\sigma_{+4}$ is indicated in the figures. }
\label{IonDist_P4N48_rc200.eps}
\end{center}
\end{figure}

It is tempting to consider that a polyelectrolyte and the ions condensed on the chain  
form a complex object.
Under some conditions, the effective charge of the complex object reverses its sign. 
This phenomenon is called charge inversion~\cite{grosberg02}.
Nguyen \textit{et al.} intended to link the phenomenon of charge inversion 
with the reentrant condensation~\cite{nguyen00}. 
They proposed that zero effective charge and charge inversion of chains are, respectively, the cause for
the condensation and for the redissolution of polyelectrolytes.
Their theory necessitates further verification either by experiments or by simulations. 
In the rest of this section, we try to calculate the effective charge of a 
polyelectrolyte, in hope of being able to clarify the relationship 
between the charge inversion and the reentrant condensation.
We simply define the complex to be the set of particles 
lying within a worm-like tube of radius $r_c$ around the chain.
This definition has been utilized in many places, for example, 
in Refs.~\cite{liu03} and \cite{netz03}. 
Usually, $r_c$ is chosen to be the distance at which the electrostatic 
attraction between a monovalent counterion and a monomer 
is equal to the kinetic energy $\frac{3}{2}k_{\rm B} T$;
in consequence, the counterion inside the tube of radius $r_c$ can not escape 
from the chain.  
We, thus, consider the total charge $Q(r_c)$ inside the tube as the effective 
charge of a chain.  In our study, $r_c$ is equal to 2.  
The results are presented in Fig.~\ref{IonDist_P4N48_rc200.eps}(a).  
Two behaviors were observed:
(1) for $\sigma_{+4}=3.0$ and 4.0, $Q(r_c)$ decreases with increasing $C_{\rm s}$,
(2) for $\sigma_{+4}\le2.0$, $Q(r_c)$ increases basically with $C_{\rm s}$. 
Notice that $Q(r_c)$ for $\sigma_{+4}=1.4$ or 2.0 attains a large positive value 
in the mid-salt region. 
Hence, $W_{\rm mm'}(r)$ should show strong repulsion in this region, 
according to the theory of Nguyen \textit{et al}. 
But, in fact, we observed a strong attractive well for $W_{\rm mm'}(r)$ and
the chains were observed to aggregate together. 
Therefore, the results do not support the theory. 

We notice that it is too simple to regard $Q(r_c)$ as the effective chain 
charge, because ion valence is not considered in the definition. 
For example, the suitable $r_c$ for tetravalent counterions should be 8.0, based upon 
the same argument to choice $r_c$ for monovalent counterions. 
In addition, $r_c=2.0$ becomes too small while $\sigma_{+4} \ge 3.0$, 
since the depletion zone around a chain for the tetravalent counterions is bigger than the tube.
Therefore, we redefine $r_c=8.0$ as the novel condensation region for tetravalent counterions, 
whereas keeping $r_c=2.0$ for monovalent counterions and for monomers on the other chains.
Considering that the condensation of coions is mainly mediated by tetravalent counterions, 
we choose $r_c=8.0$ for coions.  
The charges inside the different tubs are computed and
their sum $Q'$ is then regarded as an alternative definition 
for the effective chain charge. 
The results are presented in Fig.~\ref{IonDist_P4N48_rc200.eps}(b). 
We remark that $Q'$ looks quite differently from $Q(r_c)$. 
It shows a positive peak in the mid-salt region for $\sigma_{+4}=1.0$, 
1.4, and 2.0 but retains negative value for the other $\sigma_{+4}$.
The obtained results suggest a disconnection
between neutralization of chain charge and condensation of polyelectrolytes.

Our simulating system in the high-salt region reveals a picture more close 
to the one described by Solis~\cite{solis02}: 
although \textit{locally} overcompensated by condensed counterions, 
the net charge of a redissolved chain can be positive or negative
due to the association of coions. 
We must point out that the effective chain charge interpreted by $Q(r_c)$
or by $Q'$ in this section can be only served as a reference 
because both $Q(r_c)$ and $Q'$ are sensitive to the choice of $r_c$.
A rigorous way to study it, which can circumvent the ambiguity, is to study the electrophoretic
mobility of polyelectrolytes in salt solutions.
This part of research is currently undergoing.

\section{Conclusions}
\label{Sec_Conclusion}
We have performed molecular dynamics simulations in canonical
ensemble to investigate the properties of highly-charged 
polyelectrolytes in tetravalent salt solutions, including 
chain morphology, single-chain structure factor, swelling exponent, 
persistence length, like-charge attraction between chains and
integrated charge distribution around a polyelectrolyte. 
We have explored a wide range of salt concentration and
studied the effect of size of tetravalent counterions  
on the above properties. 

We have found that a polyelectrolyte exhibits, in turns, two conformational transitions 
upon addition of salt: a collapsed transition and a reexpanding transition,
manifested by a V-shaped curve of the square of radius of gyration 
$\langle R_{\rm g}^2 \rangle$ against salt concentration $C_{\rm s}$.
The systems at different monomer concentration $C_{\rm m}$ show similar curves.
The curves  overlap each other in the chain-reexpansion region.
In the chain-collapsed region, we have also observed that  
$A_{\rm m}(\langle R_{\rm g}^2 \rangle-\langle R_{\rm g}^{*2} \rangle)$ against $C_{\rm s}/C_{\rm m}$
lie over each other for different $C_{\rm m}$, 
where $A_{\rm m}$ and $\langle R_{\rm g}^{*2}\rangle$ weakly depends on $C_{\rm m}$.
These findings are useful in understanding the phase boundaries of the condensation 
of polyelectrolytes: on the upper boundary, $C_{\rm s}$ is roughly independent of $C_{\rm m}$
and on the lower one, $C_{\rm s}$ depends linearly on $C_{\rm m}$. 

The study of chain morphology shows that for the cases with $\sigma_{+4}$ 
(the size of the tetravalent counterions) around 1.0, the chains reveal compact and 
prolate structures 
in the vicinity of the equivalence concentration $C_{\rm s}^{*}$.
In a low-salt region and in a high-salt region, the chains show extended structures. 
At a very high salt concentration, the chains behave similarly to neutral polymers. 
For the cases with large $\sigma_{+4}$, the chains are elongated and
do not display a collapsed-transition in the mid-salt region. 
For the case with vanishing $\sigma_{+4}$, the behavior of the chains is analogous to that 
of neutral polymers as $C_{\rm s}>C_{\rm s}^{*}$.
Noticeably, our results show that $\sigma_{+4}=1.0$ is the optimal condition to pack 
polyelectrolytes into the smallest volume. 

We have shown that the single-chain structure factor $S(q)$  
of the multi-chain system in salt solutions follows the Guinier function $N(1-q^2\langle R_{\rm g}^2\rangle/3)$ 
while $q\ll R_{\rm g}^{-1}$.  
$S(q)$ then exhibits a power-law-like behavior $q^{s}$ in the region
$R_{\rm g}^{-1} \ll q \ll \sigma_{\rm m}^{-1}$,
suggesting that the scaling law, $R_{\rm g}\sim N^{\nu}$, is held for polyelectrolytes 
in salt solutions.
The swelling exponent $\nu=-1/s$ attains a minimum value 
near $C_{\rm s}^{*}$ and behaves in analogy of the morphological quantities.
It shows once more that the chains are mostly collapsed 
while $\sigma_{+4}$ is compatible with 1.0.

Our simulations incorporated explicitly the salt ions and, therefore,
were able to study the negative regime of the electrostatic 
persistence length $\ell_{\rm e}$.  This regime appears simultaneously 
with the mechanical instability of the polyelectrolytes induced by salt.   
We have examined the theory of Ariel and Andelman~\cite{ariel03b} and 
found that it fails to predict $\ell_{\rm e}$ in the \textit{so-claimed} 
valid region $b\ll \kappa^{-1} \ll (N-1)b$ but describes $\ell_{\rm e}$ 
well in the unexpected region $\kappa^{-1} \ll b$. 
Since the OSF theory is valid in the latter region,
we conjectured that the theory of Ariel and Andelman is applicable 
in the latter region.  This part deserves a detail investigation  in the future. 

We have studied like-charge attraction between polyelectrolytes 
by calculating the potential of mean force $W_{\rm mm'}(r)$ 
between monomers on different chains.
The results show that it depends strongly on salt concentration 
and ion size. 
An attractive region appears in $W_{\rm mm'}(r)$ only when 
the salt concentration is intermediate and 
the tetravalent counterions possess a size compatible with the monomers.
The most profound attractive well is happened while $\sigma_{+4}=1.0$ 
and $C_{\rm s}=C_{\rm s}^{*}$.

In the final part, the integrated charge distribution $Q(r)$ around a 
polyelectrolyte was investigated. 
We have found that the chain charge can be locally overcompensated by
condensed ions.
$Q(r)$ displays an oscillatory behavior at high salt concentrations 
while $\sigma_{+4} \ge 1.0$.
The oscillatory behavior results from a multi-layer organization of ions 
around the chain, which is in agreement with the picture predicted 
by Solis~\cite{solis02}.
We have used two different definitions to calculate the effective chain charge.
The results suggest that charge inversion is not a necessary condition to occur
reentrant condensation. 
Recently Wen and Tang have provided evidence against the connection
between charge inversion and resolubilization of polyelectrolyte 
bundles~\cite{wen04}.  
A more rigorous study  by electrophoresis performed by molecular dynamics simulations 
is undergoing.

\section{Acknowledgments}
The author acknowledges C.-L. Lee, Y.-F. Wei, and N. Pardillos for their comments. 
This material is based upon work supported by the National Science Council,
the Republic of China, under the contract No.~NSC 94-2112-M-007-023. 
Most of the simulations were running at the National Center for High-performance 
Computing. 
The author expresses his gratitude to the members and the staffs of the council and the center.

\end{document}